\documentclass[fleqn,usenatbib]{mnras}

\usepackage{txfonts}

\usepackage[T1]{fontenc}
\usepackage{ae,aecompl}

\usepackage{graphicx}	
\usepackage{arydshln}
\usepackage{color}

\newcommand{\be}{\begin{equation}}
\newcommand{\ee}{\end{equation}}
\newcommand{\rp}{R_{\rm p}}
\newcommand{\texp}{t_{\rm exp}}
\newcommand{\tov}{t_{\rm over}}
\newcommand{\sn}{{\cal S}/{\cal N}}

\def\app#1#2{%
  \mathrel{%
    \setbox0=\hbox{$#1\sim$}%
    \setbox2=\hbox{%
      \rlap{\hbox{$#1\propto$}}%
      \lower1.1\ht0\box0%
    }%
    \raise0.25\ht2\box2%
  }%
}
\def\approxprop{\mathpalette\app\relax}

\title[Exoplanetary atmosphere target selection]{Exoplanetary atmosphere target selection in the era of comparative planetology}

\author[]{J. S. Morgan,$^{1}$\thanks{E-mail: jake.morgan@postgrad.manchester.ac.uk}
E. Kerins,$^{1}$
S. Awiphan,$^{2}$
I. McDonald,$^{1}$
J. J. Hayes,$^{1}$
\newauthor S. Komonjinda,$^{3,4}$
D. Mkritchian$^{2}$
and N. Sanguansak$^{5}$
\\
$^{1}$Jodrell Bank Centre for Astrophysics, University of Manchester, Oxford Road, Manchester M13 9PL, UK\\
$^{2}$National Astronomical Research Institute of Thailand, 260 Moo 4, Donkaew, Mae Rim, Chiang Mai, 50180, Thailand\\
$^{3}$Department of Physics and Materials Science, Faculty of Science, Chiang Mai University, 239 Huay Kaew Road, Chiang Mai, \\ 50200, Thailand \\
$^{4}$Research Center in Physics and Astronomy, Faculty of Science, Chiang Mai University, 239 Huay Kaew Road, Chiang Mai, \\ 50200, Thailand \\
$^{5}$School of Physics, Institute of Science, Suranaree University of Technology, 111 University Ave., Suranaree, Nakhon Ratchasima, \\ 30000, Thailand
}

\date{Accepted XXX. Received YYY; in original form ZZZ}

\pubyear{2018}

\begin{document}
\label{firstpage}
\pagerange{\pageref{firstpage}--\pageref{lastpage}}
\maketitle

\begin{abstract}{The large number of new planets expected from wide-area transit surveys means that follow-up transmission spectroscopy studies of their atmospheres will be limited by the availability of telescope assets. We argue that telescopes covering a broad range of apertures will be required, with even 1m-class instruments providing a potentially important contribution. Survey strategies that employ automated target selection will enable robust population studies. As part of such a strategy, we propose a \textit{decision metric} to pair the best target to the most suitable telescope, and demonstrate its effectiveness even when only primary transit observables are available. Transmission spectroscopy target selection need not therefore be impeded by the bottle-neck of requiring prior follow-up observations to determine the planet mass. The decision metric can be easily deployed within a distributed heterogeneous network of telescopes equipped to undertake either broadband photometry or spectroscopy. We show how the metric can be used either to optimise the observing strategy for a given telescope (e.g. choice of filter) or to enable the selection of the best telescope to optimise the overall sample size. Our decision metric can also provide the basis for a selection function to help evaluate the statistical completeness of follow-up transmission spectroscopy datasets. Finally, we validate our metric by comparing its ranked set of targets against lists of planets that have had their atmospheres successfully probed, and against some existing prioritised exoplanet lists.}
\end{abstract}




\section{Introduction} 
\label{Intro}

The techniques of transmission photometry and spectroscopy allow astronomers to probe the atmospheres of exoplanets. By studying the variation in depth of a planet's transit curve with wavelength, individual constituents such as H$_2$O (\citealt{WASP-107Water}), Na, K and TiO (\citealt{FORSTiO}) can be identified, and the efficiency of heat redistribution from the planet's day-side to night-side can be probed (\citealt{HAT-P-32Dist}). Wider atmospheric processes such as Rayleigh scattering (\citealt{Rayleigh}), and the presence or absence of clouds and hazes (\citealt{SingHazes}, \citealt{WASP-127bSpectro}) can also be identified. This information in turn allows us to probe planetary and system dynamics, from formation theories (\citealt{FortneyCore}) to possible migration and evolution pathways as well as, in the longer term, assessing the potential habitability of planets by searching for biosignatures.

In the near future, the number of targets suitable for spectroscopic follow-up is expected to increase dramatically, with the advent of new ground and space-based transit surveys designed to target nearby hosts. These include the \textit{Next Generation Transit Survey} (\textit{NGTS} - \citealt{NGTS-1}, \citealt{NGTS-2}), the \textit{Transiting Exoplanet Survey Satellite} (\textit{TESS} - \citealt{TESS}) and, in the next decade, the \textit{PLAnetary Transits and Oscillations of stars} mission (\textit{PLATO} - \citealt{PLATO}). These surveys will provide many thousands of exoplanet candidates suitable for follow-up studies. This target-rich era will mean that we will be limited by availability of follow-up resources, rather than by a lack of of suitable candidates. It will therefore be necessary to look beyond the comparatively few large telescopes available, and bring together a more cost-effective network of smaller telescopes.

One way to view the potential contribution of small telescopes is to consider their cost-effectiveness for this type of work. \cite{SmolScopes} gives the cost-scaling of single-mirror telescopes with aperture $D$ as
\begin{equation}
\textrm{Cost} \propto D^{2.5}.
\end{equation}
For some fixed cost we can build $N$ telescopes, where $N(D) \propto D^{-2.5}.$ All other factors being equal, it is a reasonable ansatz that the resulting telescope time $T$ that a member of the community can secure on one of these telescopes will scale as
\begin{equation}
T(D) \propto N(D) \propto D^{-2.5}.
\end{equation}
The key metric is the product $TD^2$, since this dictates the time-integrated number of photons we have to work with. We might therefore expect that, through the collective effort of telescopes of aperture $D$, we can achieve a signal-to-noise ratio given by
\begin{equation}
S/N \propto \sqrt{T D^2} \propto D^{-0.25}. \label{atmoscale}
\end{equation}
The scaling in Eq.~(\ref{atmoscale}) suggests that telescopes with a broad range of apertures can make an important contribution. Of course, the scaling law ignores the fact that atmosphere studies are often limited by systematics rather than photons. It might be easier to characterise or control such systematics through the use of a single large facility rather than a number of smaller telescopes. But equally, systematic issues may be more obvious to identify from coordinated observations within a telescope network. Additionally, a clear advantage of using a number of smaller telescopes over one large one is that they can be geographically distributed to provide better global follow-up coverage of wide-area transit survey candidates. We therefore argue that future transmission spectroscopy studies should try to use all available assets, both large and small.

The Spectroscopy and Photometry of Exoplanetary Atmospheres Research Network\footnote{\url{http://www.spearnet-team.org}} (SPEARNET) is employing a globally distributed telescope network, with apertures ranging from 0.5-8m, for transmission spectroscopy studies of gas and ice giant planets. Our approach is to use an objective target selection methodology to enable completeness-corrected population studies. Our data are predominantly in the form of broadband measurements, and so multiple observations are required to characterise an atmospheric spectrum over a useful range of wavelengths. To this end, we are employing objective criteria both to select targets for observation with a given telescope asset, as well as to decide upon which filter to use to further our understanding of a given exoplanet atmosphere. Our selection strategy involves three elements: i) selection of potential targets that are observable to each telescope asset; ii) selection of targets for observation through optimal pairing of target and telescope asset; iii) assessment of how well the new observations improve characterisation of the exoplanet atmosphere. 

Stage~(i) is a ubiquitous consideration of observational astronomy. Stage~(ii) involves the application of a \textit{decision metric} to optimally pair observable targets to available telescope assets. Stage~(iii) requires consideration of data within the context of applicable planetary atmosphere models, and also feeds back to help update and inform Stage~(ii) on filter strategy. We will detail our approach to Stage~(iii) and how it feeds back into Stage~(ii) in a future paper and confine our attention here in the present paper to the description of Stage~(ii). In principle similar decision metrics could be designed for other techniques, such as emission spectroscopy (\citealt{SnellenESpec}), but our focus in this paper is on transmission spectroscopy.

The use of a decision metric for target selection has the potential advantage over expert selection of providing objective target selection criteria. This is an important attribute for future statistical studies of exoplanet atmospheres, as it enables sample completeness to be more easily assessed. Efforts to undertake such comparative exoplanetology have already begun, with studies such as \cite{FuHSTStats}, which collated \textit{Hubble Space Telescope} (\textit{HST}) spectra for 34 observed planets to search for correlations between the presence of H$_2$O spectral lines and planetary parameters. Looking ahead, \cite{ATMO} gathered together 117 planets identified as observationally significant for inclusion in their grid of forward atmospheric transmission spectra, generated by the ATMO modelling suite. Population studies are vital to understand the variety of exoplanet compositions and its dependence on planet and host characteristics, as well as to provide a probe of planet formation and dynamics. As the list of potential targets grows rapidly over the coming decade, these types of studies will clearly become increasingly common and more sensitive.

Our paper is structured as follows: Section \ref{DecMet} details the derivation of our decision metric that is employed in Stage~(ii) of the SPEARNET strategy. Section \ref{MetExtensions} explores the various extensions that can be employed for specific scenarios, as well as the scaling of the metric upon various physical parameters and the implications of this. Section~\ref{Apply} then demonstrates how the metric can be used to select telescope assets and optimize observation strategy. It also validates the metric against previously-studied samples. We summarise our work in Section~\ref{Conclude}.

\section{The Decision Metric} 
\label{DecMet}

We base our decision metric upon the contribution to the signal-to-noise ratio during transit of an exoplanet of radius $\rp$, that possesses an atmosphere of scale height $H \ll \rp$ with some non-zero opacity at the observation wavelength. 

In the ideal case, the noise is fundamentally limited by Poisson noise from the host. In Section~\ref{moreNoise} we consider the effect of some additional sources of noise not included within our default decision metric, such as sky and transit baseline noise, red noise and atmospheric scintillation. However, we neglect several other potentially important sources of noise. For example, we stick to the ``quiet star'' approximation, in which the contribution to the overall noise from stellar activity is negligible. We also treat the host as a uniform disk.

Some of these neglected factors may well be relevant to the success or failure of observations of a given target with a given facility. Clearly, a decision metric cannot encode all potentially relevant information without becoming so complex that it is rendered unusable. However, it can encode basic planet and host characteristics that are most favourable for transmission spectroscopy studies with a given facility. Whether a metric ultimately encodes sufficient information for it to be useful in practise can be tested by comparing it with samples of previously successfully studied systems, which we will do in Section~\ref{compare}.

\subsection{Derivation}
\label{define_met}

On discovery of a transit candidate, the key host and planet parameters, except for $H$, are typically measured or inferred. We can substitute $H$ in the case of an atmosphere that can be approximated as an ideal gas, existing in an annulus of area
\begin{equation}
A_{\rm ann} = 2\pi \, R_{\rm p} H ,
\label{A_ann}
\end{equation}
where
\begin{equation}
H = \frac{k_{\rm B} T_{\rm p}}{\mu g} \propto \frac{T_{\rm eq}}{\mu g} \propto \frac{T_{\rm eq} R_{\rm p}^2}{\mu M_{\rm p}},
\label{gas}
\end{equation}
with $k_{\rm B}$ representing the Boltzmann constant, $T_{\rm p}$ the temperature of the planetary atmosphere, $\mu$ the atmospheric mean molecular weight, $M_{\rm p}$ the planet mass and $g$ the planet surface gravity. Whilst $T_{\rm p}$ may not be known, the inferred equilibrium temperature, $T_{\rm eq}$, provides a reasonable proxy. For the purposes of our comparison, we treat $\mu$ as fixed.

The planet's surface gravity may not be determinable without additional measurements such as radial velocity or timing variations, in which case the mass must be substituted through a mass--radius relation. In Section~\ref{KnownMass} we develop a suitable metric for the case where both the planet mass and radius are known. In the remainder of this section we pursue a suitable metric for the case where only the radius is known, as will be the case for transit detections prior to follow-up confirmation. We will test its performance against the metric that includes the planet mass in Section~\ref{KnownMass}. Naturally, it is always better to include mass information when available. However, follow-up observations aimed at obtaining the masses of candidates from wide-area surveys such as \textit{NGTS}, \textit{TESS} and \textit{PLATO} will take time, and therefore present a potential bottle-neck to the progress of transmission spectroscopy surveys. An effective decision metric that relies only on primary transit observables can potentially circumvent this bottle-neck and is therefore worth pursuing.

\begin{figure}
	\includegraphics[width=\columnwidth]{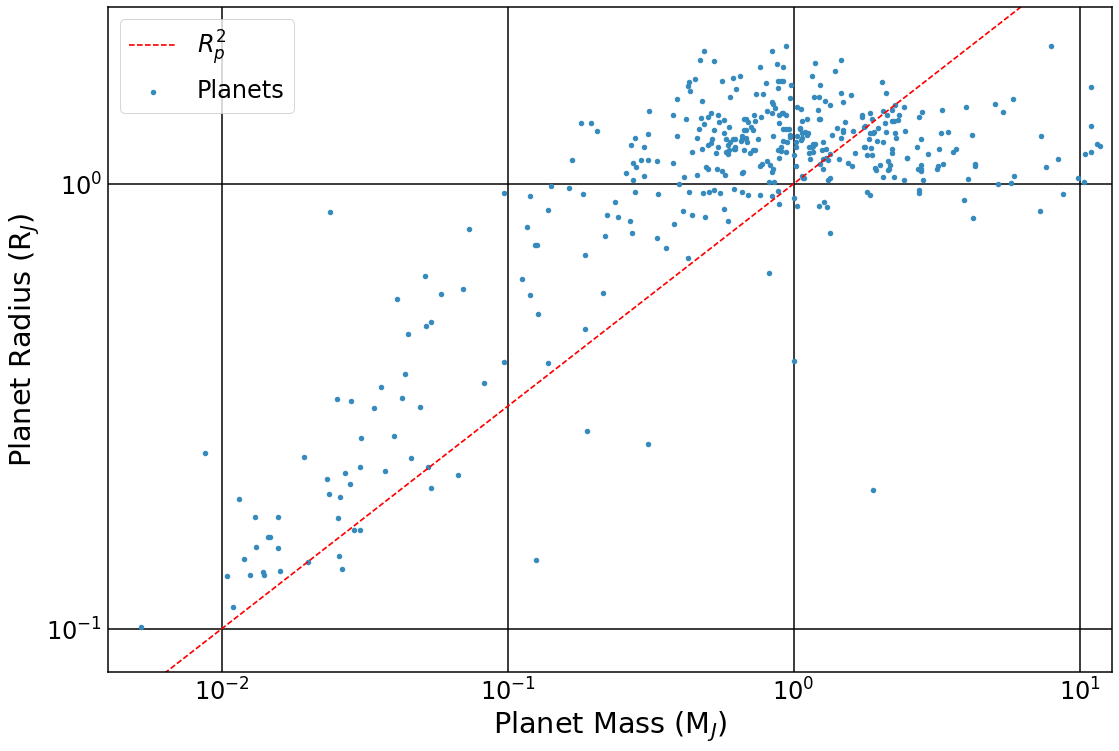}
	\caption{The mass-radius relation for detected exoplanets, where both mass and radius are known to within 20\%. Below around $0.3\,M_{\rm J}$ (or around $0.8\,R_{\rm J}$) the distribution is seen to follow an approximate $M_{\rm p} \propto \rp^2$ relation (red line), though with significant scatter. Above these limits the relationship is flat, though again with significant scatter. }
	\label{MRPlot}
\end{figure}

The mass--radius ($M_{\rm p} - \rp$) relation is often expressed as a simple power law, $M_{\rm p} \propto \rp^n$. For example, \cite{Zeng16} find that $n \simeq 3.7$ for rocky planets. However, generally the relation depends upon several factors including the planet composition, equation of state, age, tidal effects and host irradiation. Indicative values for larger planets based upon detections so far suggest $n \sim 2$ for Earth- to Saturn-mass planets and $n \sim 0$ for more massive planets \citep[e.g.][]{Bashi}. Figure \ref{MRPlot} is a scatter plot using data from the Extrasolar Planets Encyclopaedia database\footnote{\url{http://exoplanets.eu/}.} exploring this relation. 

Current data on large exoplanets show significant scatter, and therefore no tight correlation between $M_{\rm p}$ and $\rp$.  For the purposes of developing a useful metric, we will show that the best approach for larger planets ($R > 0.8\, R_{\rm J}$) is to treat $M_{\rm p}$ and $\rp$ as independent (i.e. $n = 0$). For smaller planets we will show in Section~\ref{KnownMass} that a value of $n = 2$ is preferred. For the following derivations we shall therefore explicitly include $n$ within our metric. 

Applying our scaling for surface gravity to Eq. (\ref{gas}) and dividing out constants, we recover a scaling for $H$ of
\be
H \approxprop T_{\rm eq} \rp^{2-n} .
\label{hscale}
\ee
We employ the \textit{maximal atmospheric signal} limit, corresponding to the number of host photons that are blocked by a completely opaque atmosphere (in addition to those blocked by the body of the planet) during transit. In a single exposure of duration $t_{\rm exp}$, the photon count from a host of magnitude $m_*$ at a telescope with magnitude zero-point $m_{\rm zp}$ at the observation wavelength is

\begin{equation}
n_* = 10^{0.4(m_{\rm zp} - m_*)} t_{\rm exp},
\label{NSingle}
\end{equation}

\noindent where $t_{\rm exp}$ must be expressed in seconds when using the standard definition of $m_{\rm zp}$ as the source magnitude required for a detection rate of 1 photon/sec. This gives a maximal atmospheric signal

\begin{eqnarray}
{\cal S}_* &=& n_*\frac{A_{\rm ann}}{\pi R_*^2}
\label{FirstSig}
\\
&=& 10^{0.4(m_{\rm zp} - m_*)} \texp \frac{2\,R_{\rm p} H}{R_*^2} ,
\label{AtmTotal}
\end{eqnarray}

\noindent where $R_*$ is the radius of the host. 

For a uniform disk, the transit depth $\delta$ is given by
\begin{equation}
\delta = \frac{R_{\rm p}^2}{R_*^2}.
\label{Depth}
\end{equation}
Employing this in the form $R_{\rm *}^{-2} = R_{\rm p}^{-2}\delta$ and using our proportionality scaling from Eq.~(\ref{hscale}), we find the signal per exposure to be
\be
{\cal S}_* \propto 10^{0.4(m_{\rm zp} - m_*)} \texp T_{\rm eq} \rp^{1-n}  \delta. 
\label{SNOne}
\ee
In the most optimistic situation the signal-to-noise ratio per exposure will fundamentally be limited by the Poisson noise from the host flux not blocked out during transit:
\be 
\varepsilon_* = [n_*(1-\delta)]^{1/2} \propto 10^{0.2(m_{\rm zp} - m_*)} \texp^{1/2} (1- \delta)^{1/2}.
\label{nseOne}
\ee
Bringing together Eqs. (\ref{SNOne}) and (\ref{nseOne}) gives a signal-to-noise ratio per exposure of
\be
{\cal S}/{\cal N}_* \equiv {\cal S}_*/\varepsilon_* \propto 10^{0.2(m_{\rm zp} - m_*)} \texp^{1/2} T_{\rm eq} \rp^{1-n}  \delta (1- \delta)^{-1/2}. \label{SNratOne}
\ee

The ability to fit a transit model will depend not just on the quality of each observation but also on the number of exposures obtained during the transit. Our metric must therefore reflect this fact. Let us consider a uniform time series of exposures collected during a transit of duration $t_{14}$. The total signal-to-noise ratio summed over all $n_{\exp}$ exposures will be

\begin{eqnarray}
{\cal S}/{\cal N}_{*,\rm tot} & = & ({\cal S}/{\cal N}_*) n_{\rm exp}^{1/2} = {\cal S}/{\cal N}_* \left( \frac{t_{14}}{\texp + \tov} \right)^{1/2} \nonumber \\
 & \propto & 10^{0.2(m_{\rm zp} - m_*)}\frac{T_{\rm eq} \rp^{1-n}  \delta}{ (1- \delta)^{1/2}} \left( 
 \frac{t_{14} \texp}{\texp + \tov} \right)^{1/2},
\label{SNAll}
\end{eqnarray}

\noindent with $\tov$ being the duration between the end of one exposure and start of the next, e.g. due to the detector read-out overhead. 
In a regime where $(R_{\rm p} + H) \ll R_*$, we can safely approximate $(1-\delta)^{-1/2}$ as unity, though this factor may remain relevant for scenarios with hot-Jupiters around M-dwarfs, or in instances where the planet is undergoing atmospheric escape \citep{GJ436ExoEscape}.

We can therefore define the \textit{decision metric}, ${\cal D} \propto {\cal S}/{\cal N}_*$ as
\be
{\cal D} = C_{\rm T}\, 10^{-0.2 m_*} t_{14}^{1/2} T_{\rm eq} \left(\frac{\rp}{0.8\,R_{\rm J}}\right)^{1-n}  \delta \quad (\rp \ll R_*),
\label{OldMetric}
\ee
where
\be
C_{\rm T} = N_\lambda^{-1/2} 10^{0.2 m_{\rm zp}}\left( \frac{\texp}{\texp + \tov} \right)^{1/2}
\label{TelFac}
\ee
contains all of the telescope-dependent factors. The factor $N_{\lambda}$ in Eq. \ref{TelFac} specifies the number of spectral bins spanning the observed wavelength range. It is included to allow both for photometric observations, where $N_{\lambda} = 1$, or for spectroscopic observations where $N_{\lambda} \gg 1$. We will consider both photometric and spectroscopic modes of observation in the following sections.

Note that in Eq. \ref{OldMetric} we choose to normalise the planet radius to a value of $0.8\, R_{\rm J}$. We will show in Section~\ref{KnownMass} that, in the absence of planet mass measurements, it is best to use different values for $n$ above and below this radius, as indicated by the mass-radius distribution plotted in Figure~\ref{MRPlot}. Normalising $R_{\rm p}$ to $0.8\, R_{\rm J}$ is therefore necessary to ensure continuity of metric scores across this boundary.

Eq~(\ref{OldMetric}) forms the basis for Stage~(ii) of the three-stage procedure employed by the SPEARNET team, as outlined in Section~\ref{Intro}. It is important to stress that Eq~(\ref{OldMetric}) solves the problem of the optimal assignment of exoplanet targets to telescope assets, based on overall signal-to-noise considerations. It does not by itself indicate how easily a planet atmosphere can be characterised. Neither can we achieve this through summed combinations of metrics determined for different wavelengths. The reason for this is that the metric by itself does not contain any information \textit{a priori} about the wavelength-dependence of the planet's atmospheric opacity. To this end, SPEARNET is employing a third stage that compares new data to atmosphere models in order to assess the resulting improvement in characterisation. A detailed discussion of Stage~(iii) will be the subject of a forthcoming paper.

\subsection{Additional Noise Considerations} 
\label{moreNoise}

The metric defined by Eq.~(\ref{OldMetric}) can be extended further if desired, to include additional sources of noise. One such source is the uncertainty in the out-of-transit baseline. Assuming a uniform set of exposures obtained over a period of $t_{\rm base}$ out of transit (we shall assume one hour of observations both before and after the transit, for two hours total duration), the precision with which the baseline can be determined is given by

\be
\varepsilon_{\rm base} = \varepsilon_* n_{\rm base}^{-1/2} = \varepsilon_* \left( \frac{\texp + \tov}{t_{\rm base}} \right)^{1/2} ,
\ee
   
\noindent where $n_{\rm base}$ is the number of baseline epochs observed. We can include this within our error budget by defining

\be
\varepsilon_*' = \sqrt{\varepsilon_*^2 + \varepsilon_{\rm base}^2} = \sqrt{\varepsilon_*^2 \left[1 + \left(\frac{t_{14}}{t_{\rm base}} \right)\right]} , \label{baseNoise}
\ee

\noindent and making the substitution $\varepsilon_* \rightarrow \varepsilon_*'$ in Eq.~(\ref{SNratOne}). Typically, $t_{14} \approx t_{\rm base}$, yielding an overall correction of $\sqrt{2}\varepsilon_*$.

Another potential source of noise arises from the sky background

\begin{equation}
n_{\rm sky} = 10^{0.4(m_{\rm zp}-m_{\rm sky})}t_{\rm exp},
\end{equation}

\noindent where $m_{\rm sky}$ is the apparent magnitude contribution of the sky within the observation point spread function (PSF). For example, taking a PSF of radius of $\Theta_{\rm PSF}$, $m_{\rm sky}$ is given by

\begin{equation}
	m_{\rm sky} = -2.5\log\left[\pi \Theta_{\rm PSF}^2\right]+\mu_{\rm sky}.
\end{equation}

\noindent Here $\mu_{\rm sky}$ is the wavelength-dependent sky brightness expressed in magnitudes per unit angular sky area. Both $\mu_{\rm sky}$ and $\Theta_{\rm PSF}$ are dependent on the chosen observing site. $\Theta_{\rm PSF}$ will be set by the seeing scale for focussed observations, or it may be much larger for defocussed observations.

For spectrographs using a slit of width $a$, the magnitude contribution will be

\begin{equation}
	m_{\rm sky} = -2.5\log\left(2\,a\Theta_{\rm PSF}\right)+\mu_{\rm sky}.
\end{equation}

The corresponding sky noise contribution is therefore

\be
\varepsilon_{\rm sky}=10^{0.2(m_{\rm zp}-m_{\rm sky})}t_{\rm exp}^{1/2} ,
\ee

\noindent which can be incorporated into Eq.~(\ref{SNratOne}) by the substitution $\varepsilon_* \rightarrow (\varepsilon_*^2 + \varepsilon_{\rm sky}^2)^{1/2}$. 

A third noise source arises from scintillation, which causes brightness fluctuations due to turbulence in the Earth's atmosphere, e.g. \cite{OsbornScint}. We model this noise contribution using the modified form of Young's approximation which those authors set out, which is of the form
	
\begin{equation}
\varepsilon_{\rm scint}^2 = 0.0032C_{\rm Y}D^{-2/3}t_{\rm exp}^{1/2}e^{-h_{\rm obs}/8000} ,
\end{equation}
	
\noindent where $C_{\rm Y}$ is an empirical coefficient, $D$ is the telescope diameter and $h_{\rm obs}$ is the observatory altitude in metres (we assume $h_{\rm obs} = 2300$~m). Values of $C_{\rm Y}$ are listed for several sites in \cite{OsbornScint}; for the sake of consistency, we use the median values and set the parameter as 1.5 for non-listed sites.

\begin{figure*}
	\includegraphics[width=\columnwidth]{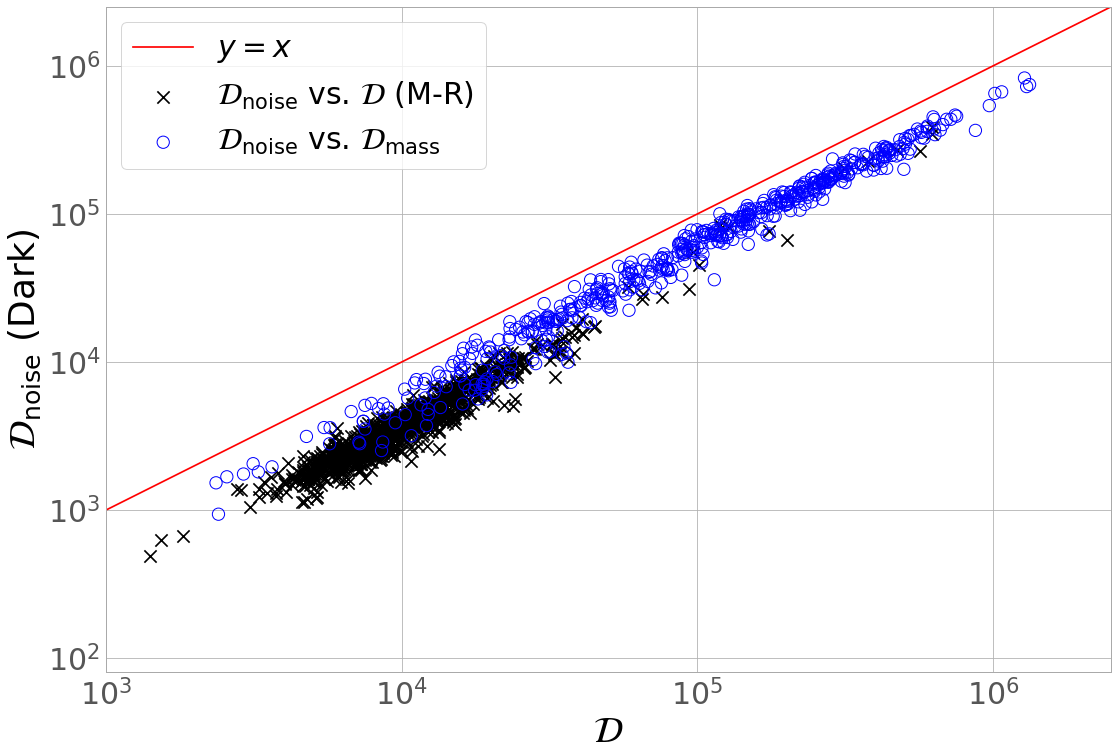}
	\includegraphics[width=\columnwidth]{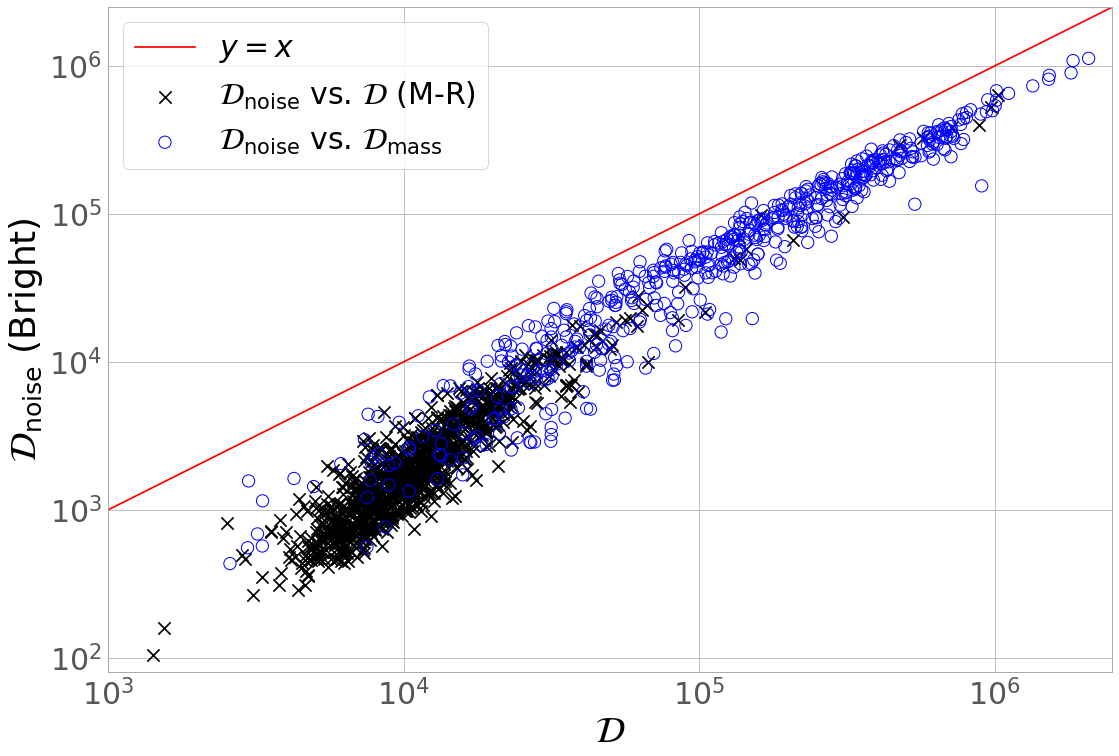}
	\includegraphics[width=\columnwidth]{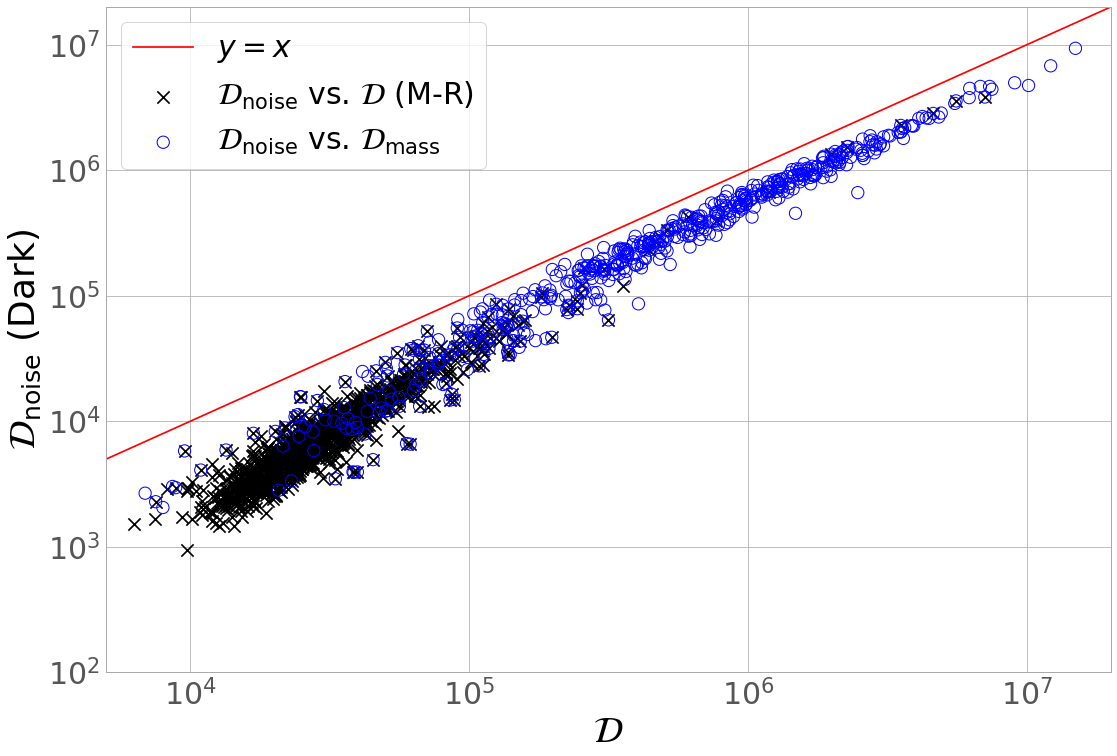}
	\includegraphics[width=\columnwidth]{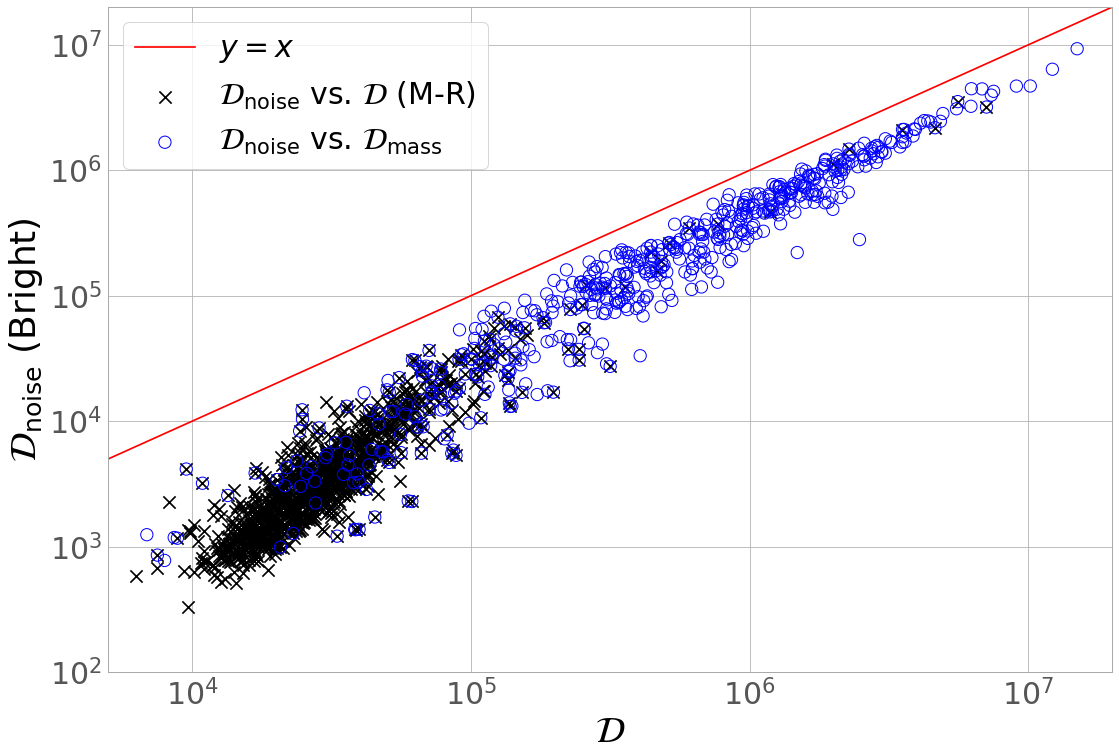}
	\caption{The effect of various noise terms on the metric. The upper two panels shows the metric score incorporating sky, baseline, scintillation and red noise terms, ${\cal D}_{\rm noise}$, compared to the score when these sources are neglected, ${\cal D}$, assuming no defocusing. The top-left panel uses a dark sky background of $R=21.2~\textrm{mag/arcsec}^2$ and an aperture of $3''$ radius, while the top-right uses a bright sky background of $R=17.9~\textrm{mag/arcsec}^2$ and a $5''$ radius aperture. The lower panels show the same comparison for a typical defocused photometry scenario with a defocus aperture of $14''$ diameter, with sky backgrounds of $r'=19.8$ and $r'=17.5$ respectively. The observing altitude $h_{\rm obs}$ is set at 2300m. Sky noise has a larger effect on the metric for the defocused case and for low-ranked targets, but in all cases it is clear that a tight correlation is retained with our simple metric presented in Eq. (\ref{OldMetric}) for higher-ranking planets, which would be selected as the most suitable observational targets. The sample consists of 1558 planets, split between those with known masses (blue circles, using Eq. \ref{MetMass} in Section \ref{KnownMass}) and those without (black crosses), using Eq. (\ref{OldMetric}).}
	\label{ExtraTermsPlot}
\end{figure*}	

Finally, we include a fixed red noise budget of $\varepsilon_{\rm RN} = 100$ ppm, to account for systematic effects. Although we cannot model this in detail for every possible telescope set-up, we can nonetheless include it as a noise floor, which will depend only weakly on the total number of exposures. With these contributions, our total noise budget becomes

\begin{equation}
\varepsilon_{\rm tot} = \sqrt{\varepsilon_*^2 + \varepsilon_{\rm base}^2 + \varepsilon_{\rm scint}^2 + \varepsilon_{\rm RN}^2} .
\end{equation}

\begin{figure*}
	\includegraphics[width=\columnwidth]{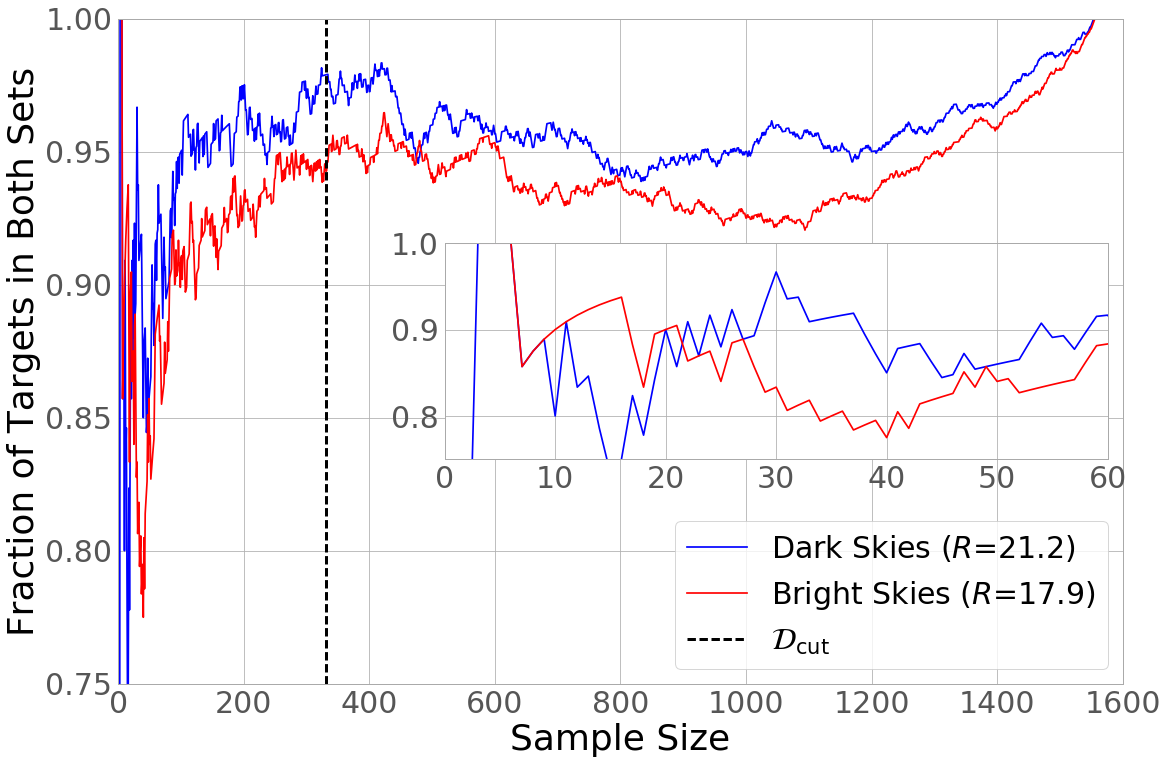}
	\includegraphics[width=\columnwidth]{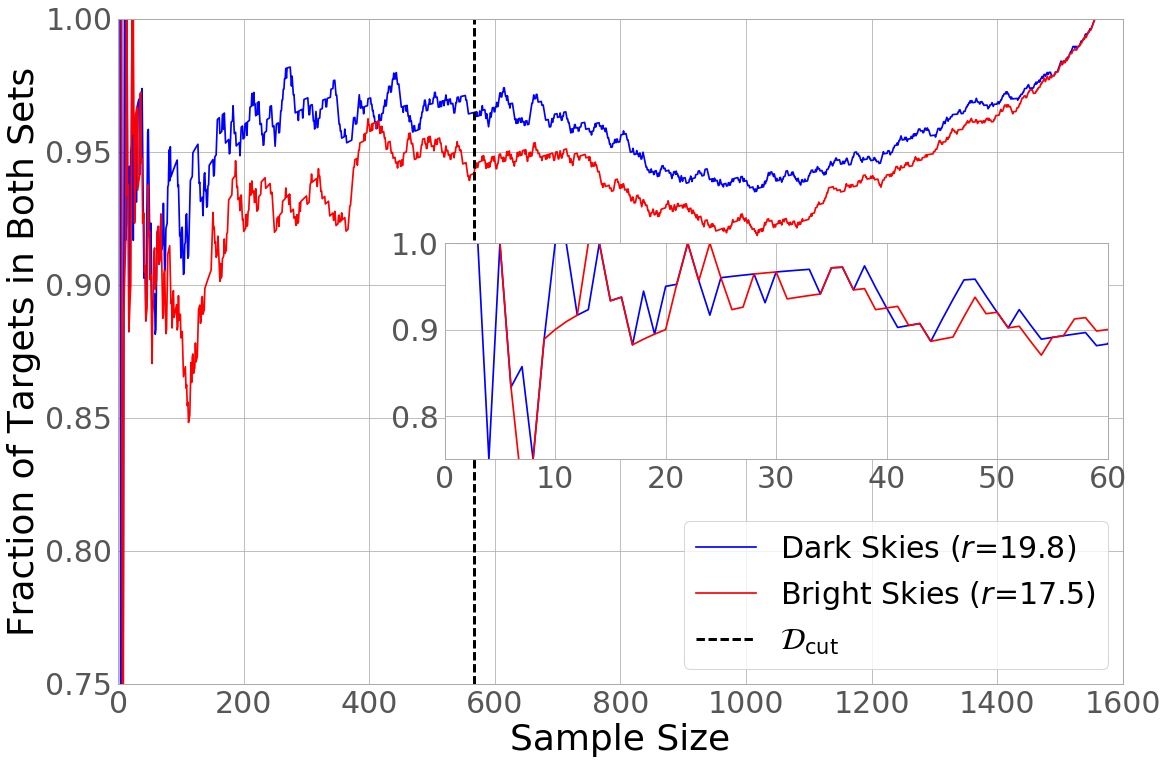}
	\caption{Fractional overlap between the two ranked lists (with and without noise budget) for the four samples in Figure \ref{ExtraTermsPlot}. The left panel shows focused observations, while the right is for defocused. $\cal D_{\rm cut}$ denotes the cut-off for selecting viable targets for the purposes of comparing different set-ups (Section \ref{CutoffMade}). The inset zoomed panels in each plot show the overlap for the top 60 candidates.}
	\label{CutOverlaps}
\end{figure*}

Figure~\ref{ExtraTermsPlot} illustrates the effects of including these additional terms, for both dark and bright skies. Although individual scores are suppressed by modelling noise terms, the overall distribution holds. Under bright skies (right panels), the $\cal D$ scores of the lowest-ranked targets are suppressed by an order of magnitude, as the sky background flux becomes comparable to the source flux for these generally faint objects. Sky backgrounds should not be neglected for these cases. However, the highest-ranked targets display only minor scatter; introducing noise terms does not change their rank order significantly, meaning the same top targets are selected in both cases. It is these planets we wish to identify with our ranking scheme.

Figure \ref{CutOverlaps} quantifies this. For focused observations under dark skies (blue line in left panel of Figure \ref{CutOverlaps}) selecting the top 331 targets (black vertical dashed line), 324 appear in both sets, giving a 97.8\% overlap. Under bright skies, the overlap is 95\%. We can say that for this sample of top targets, neglecting noise terms induces only small changes in the ranking order (typically of only a few places), and so has only a minor effect on the composition of the overall sample returned. As long as the chosen cut-off for this top sample is sensible, our improved noise scheme is not necessary to recover this desired set of targets. Given this and that some terms in our noise scheme are poorly quantified, e.g. scintillation and red noise, the signal-to-noise scaling in Eq.~(\ref{SNAll}) is sufficient for our purposes. We adopt this form throughout the remainder of the paper. We describe the process of making a viability cut in Section \ref{CutoffMade}.

\section{Metric Extensions and Parameter Bias}
\label{MetExtensions}

\subsection{Extension for Known Planetary Masses}
\label{KnownMass}

\begin{figure*}
	\includegraphics[width=\columnwidth]{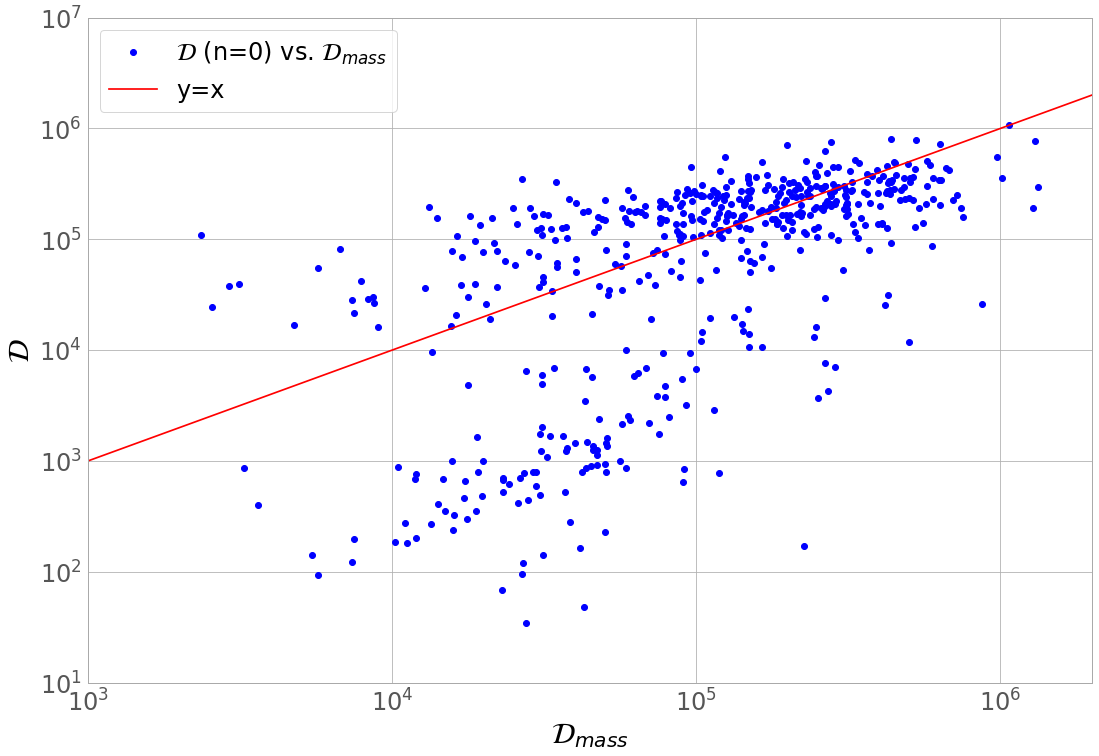}
	\includegraphics[width=\columnwidth]{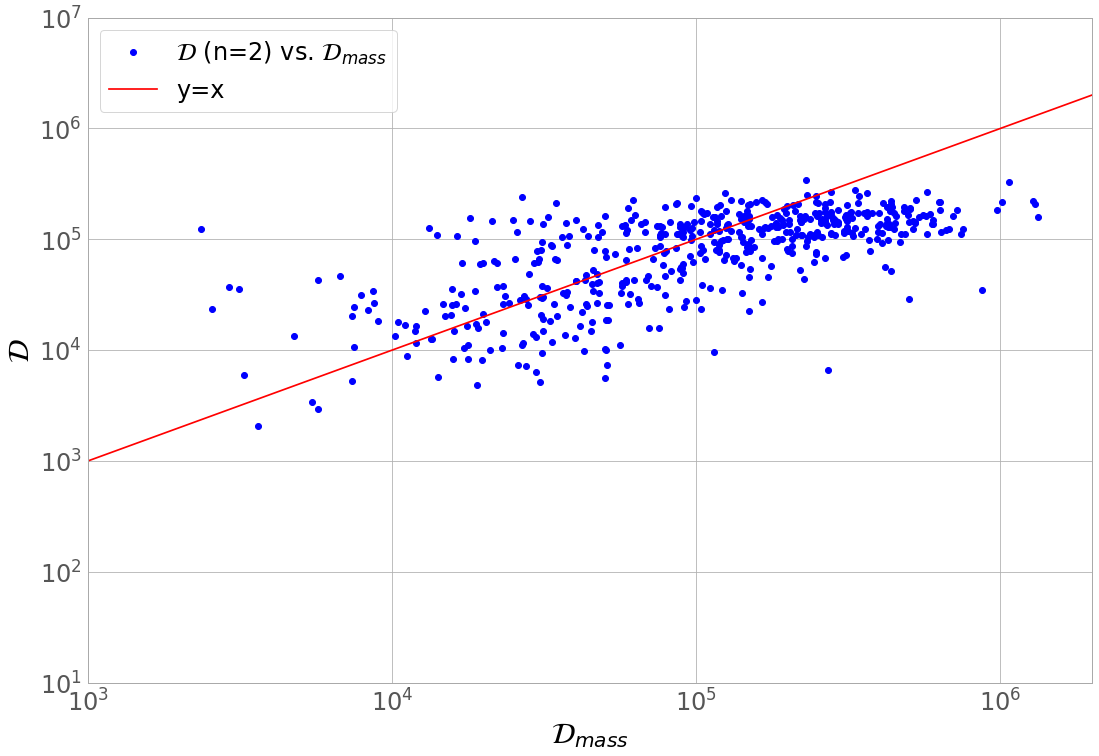}
	\includegraphics[width=\columnwidth]{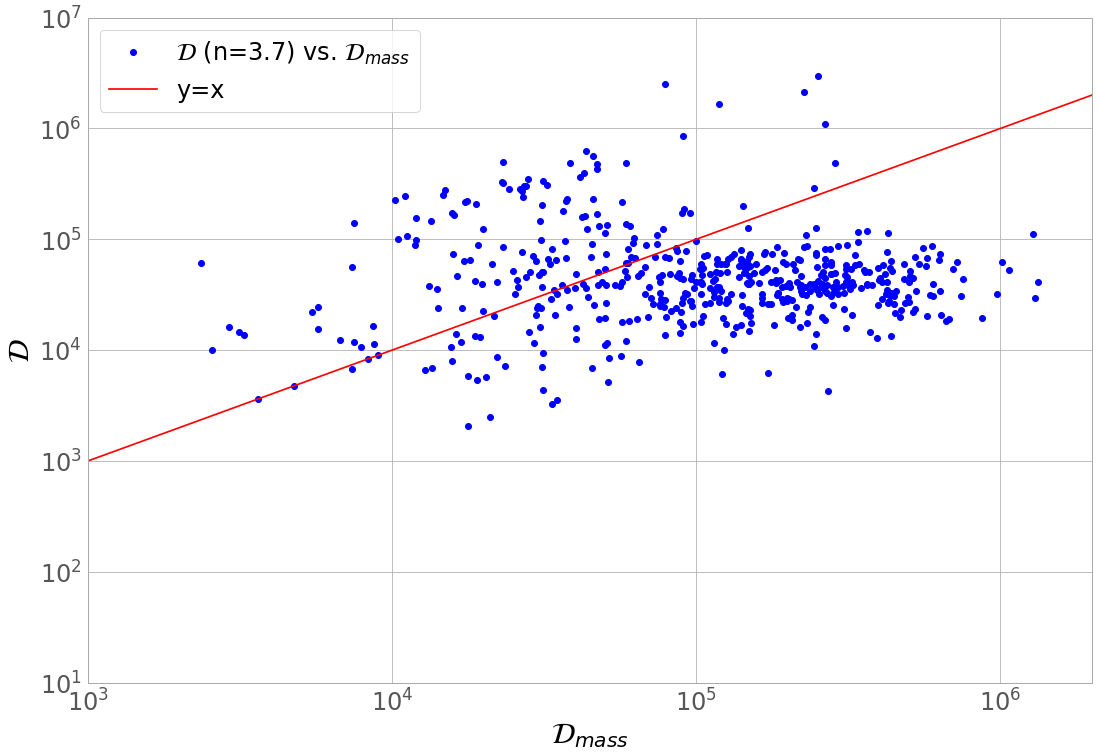}
	\includegraphics[width=\columnwidth]{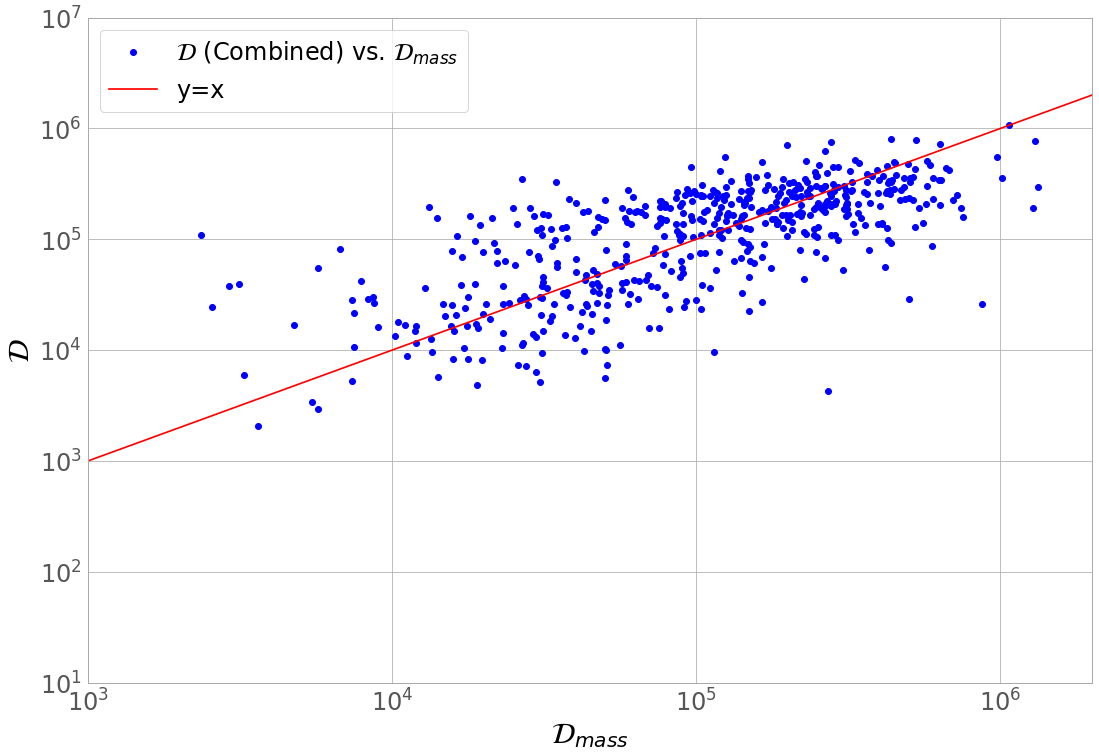}
	\caption{The correlation between the metric, ${\cal D}$, employing only the planet radius, to ${\cal D}_{\rm mass}$ from Eq. (\ref{MetMass}) that additionally uses the planet mass. Metric scores are shown for a sample of 527 planets with well-determined mass and radius. ${\cal D}$ is evaluated for three different $M_{\rm p} \propto R_{\rm p}^n$ relations discussed in the text: i) $n=0$ (top-left); ii) $n=2$ (top-right); and iii) $n=3.7$ (bottom-left). These values may be appropriate for gas giants, super-Earths/mini-Neptunes and Earth-mass planets respectively. It is clear that no single value for $n$ provides a good overall correlation, with $n = 0$ providing the best approximation for the gas giant regime and $n = 2$ for lower mass planets. The bottom-right panel shows the result of using $n = 2$ for $R < 0.8 R_{\rm J}$ and $n = 0$ above this, providing a more satisfactory overall correlation. It is clearly preferable to employ ${\cal D}_{\rm mass}$ when the planet mass is known, but using ${\cal D}$ with this adopted two-part $M-R$ relation can form a suitable substitute when the planet mass is unknown or unreliable.}
	\label{MassEffect}
	
\end{figure*}
As noted in Section \ref{define_met}, the mass of a given transiting planet must be recovered from follow-up observations. As more such radial velocity follow-up observations are undertaken, particularly of new \textit{TESS} targets (\citealt{CloutierRV}), the subset of transiting planets with known masses will become large, to the point where statistically significant samples can be recovered without the need for a mass substitution. We can return to Eqs. (\ref{gas}) and (\ref{AtmTotal}) and apply our scaling for the surface gravity $g$ using the mass directly, yielding

\be
{\cal S}  \propto 2\times 10^{0.4(m_{\rm zp}(\lambda) - m_*(\lambda))} t_{14}  \left( \frac{\texp}{\texp + \tov} \right) \frac{T_{\rm eq} \rp  \delta}{M_{\rm p}}. 
\label{SNmass}
\ee

Our overall metric using known planet masses will then be

\begin{equation}
{\cal D_{\rm mass}} = C_{\rm T}(\lambda) 10^{-0.2m_*} t_{14} \frac{T_{\rm eq} \rp \delta}{M_{\rm p}} .
\label{MetMass}
\end{equation}

Figure~\ref{MassEffect} compares the $\cal D$ and ${\cal D}_{\rm mass}$ scores, evaluated from Eqs. (\ref{OldMetric}) and (\ref{MetMass}) respectively, of a sample of 527 planets with well-determined masses. Three of the panels show the comparison for three single values of $n$ discussed in Section~\ref{define_met} ($n = 0, 2$ and 3.7). As might be expected, all three relations yield considerable scatter and systematic offsets between ${\cal D}$ and ${\cal D}_{\rm mass}$, with no single value of $n$ providing a satisfactory correlation. 

The last panel of Figure~\ref{MassEffect} shows the result of adopting a piece-wise $M_{\rm p} - R_{\rm p}$ relation with $n = 2$ for $R < 0.8\, R_{\rm J}$ and $n=0$ above this. This time ${\cal D}$ and ${\cal D}_{\rm mass}$ exhibit a reasonable correlation. Whilst there remains a significant level of scatter between the two metrics, ${\cal D}$ represents a viable alternative to ${\cal D}_{\rm mass}$ for cases where the planet mass is not known. It is clearly best to use ${\cal D}_{\rm mass}$ where mass information is available, but we conclude that ${\cal D}$ represents a workable proxy otherwise. 

For the remainder of this paper we adopt the same two-piece $M_{\rm p} - R_{\rm p}$ relation when evaluating decision metrics that have no explicit mass dependence. We use this relation to handle targets that have no reliable mass estimate. For planets that do have such estimates, we employ the extension detailed in Eq. (\ref{MetMass}).

\subsection{Extension for Long-Term Studies}
\label{Extend}

\begin{figure*}
	\includegraphics[width=\columnwidth]{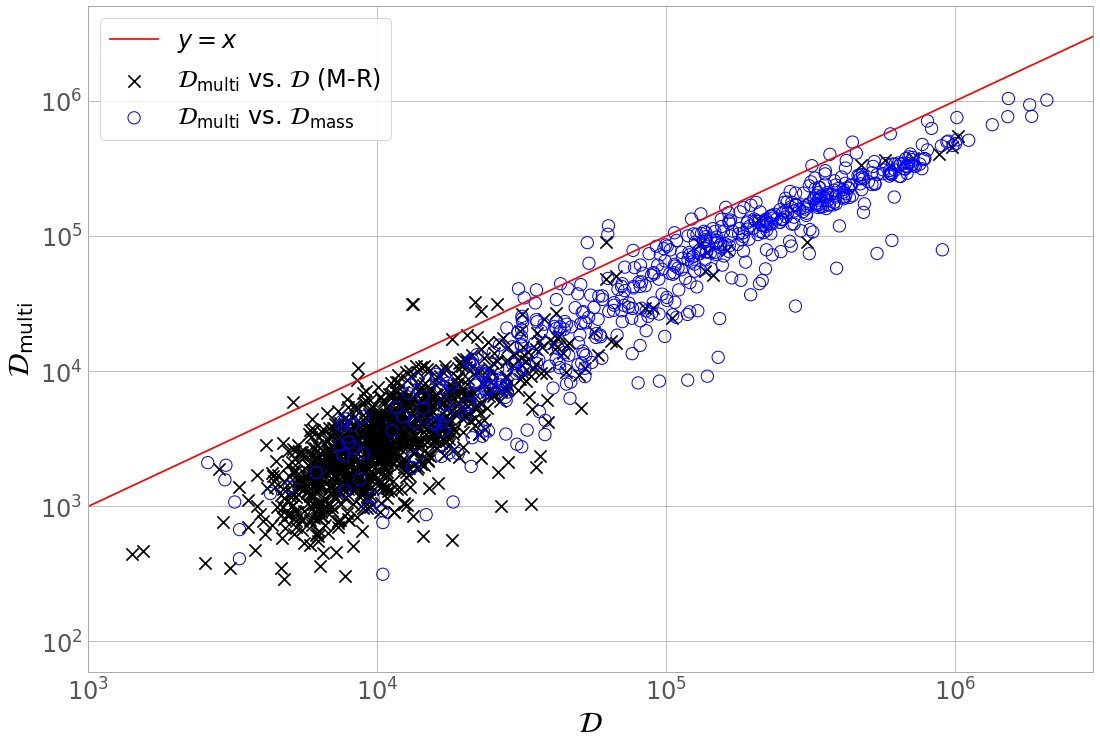}
	\caption{Comparison between the standard metric, ${\cal D}$, from Eq. (\ref{OldMetric}) and that applicable for multi-transit observations, ${\cal D}_{\rm multi}$ (Eq~\ref{multi}). Targets with known masses instead use ${\cal D_{\rm mass}}$ (Eq. \ref{MetMass}) for comparison to ${\cal D}_{\rm multi}$. Planets appearing above the ${\cal D} = {\cal D}_{\rm multi}$ line (in red) all have $P < 1$ day. There is a reasonably strong correlation between $D$ and $D_{\rm multi}$ for high ranking planets, implying that the choice between the two metrics is not crucial in this regime.}
	\label{MultiEffect}
\end{figure*}

The metric in Eq.~(\ref{OldMetric}) is appropriate for observations carried out over a single transit. Campaigns using smaller telescopes may well require multiple observations in order to bring about sufficient improvement in signal to noise through co-adding observations. In this case one can modify ${\cal D}$ to include the co-added improvement resulting from observations over a number of transits $N_{\rm trans}$, spanning some fixed total observing time. 

This leads to a modified multi-transit decision metric, $D_{\rm multi} \propto N_{\rm trans}^{1/2} \propto P^{-1/2}$, which can be defined in terms of ${\cal D}$ as
\be
{\cal D}_{\rm multi} = {\cal D} \left( \frac{P}{\mbox{days}}\right)^{-1/2}, \label{multi} 
\ee
where $P$ is the planet orbital period in days. ${\cal D}_{\rm multi}$ preferentially favours shorter-period planets, which will be observable more often in a campaign of fixed total duration. The multi-transit decision metric is suitable only for periods that do not alias with the Earth's rotation period, otherwise only a fraction of possible planetary transits may be visible at night from a given location. Hence, rankings returned for such cases must be filtered for time and location to correct for this.

Figure \ref{MultiEffect} compares ${\cal D}$ and ${\cal D}_{\rm multi}$, evaluated from Eqs. (\ref{OldMetric}), (\ref{MetMass}) and (\ref{multi}). We can see that, whilst ${\cal D}_{\rm multi}$ scores are systematically lower than those for ${\cal D}$, this is simply because we have normalised $P$ in Eq~(\ref{multi}) to days and most of the sample comprises planets with $P > 1$~day. It is the correlation between ${\cal D}$ and ${\cal D}_{\rm multi}$ that determines differences in planet rank ordering, and this remains quite tight, especially for higher ranked systems. This implies that ${\cal D}_{\rm multi}$ provides a modest, though not necessarily essential, correction to ${\cal D}$.

\subsection{Dependence on Host and Planet Properties} 
\label{bias}

If our decision metric is to prove useful for future statistical studies of exoplanetary atmospheres, it is important that it should not depend too sensitively on specific physical parameters of the host or planet, to the extent that it selects a subset of systems with properties that span a very narrow range compared to those that could be probed. Here we look at how our decision metric scales with underlying planet and host parameters.

In the usual limit of $\rp \ll R_*$, we see from Eq.~(\ref{OldMetric}) that

\be
{\cal D} \approxprop (F_* t_{14} )^{0.5} T_{\rm eq} \rp^{1-n} \delta, \label{metscl} 
\ee
where $F_*$ is the host flux. Reducing each term in Eq.~(\ref{metscl}), we note that $F_* \propto L_* d^{-2}$, where $L_*$ is the host luminosity and $d$ the distance of the host from the observer. This is only strictly true when considering the bolometric spectral energy distribution (SED), but is still useful for examining approximate dependencies. From Newton's laws of motion, the transit duration $t_{14} \propto M_*^{-0.5} R_* a^{0.5}$, where $a$ is the semi-major axis of the planet's orbit. For the third and fourth terms in Eq.~(\ref{metscl}), we have $T_{\rm eq} \propto L_*^{0.25} a^{-0.5}$ and $\delta = (\rp / R_*)^2$. Gathering all of these dependencies together gives

\be
{\cal D} \approxprop L_*^{0.75} M_*^{-0.25} R_*^{-1.5} \rp^{(3-n)} a^{-0.25} d^{-1} . 
\label{metfund}
\ee

We can use simple power-law, mass--luminosity and mass--radius relationships for main--sequence stars (specifically $L_* \propto M_*^x$ and $R_* \propto M_*^y$), to contract Eq.~(\ref{metfund}). For stars between 0.4 and 2~M$_{\odot}$, $x \simeq 4$ and $y \simeq 0.8$, whilst for more massive stars of up to 20~M$_{\odot}$, $x \simeq 3.5$ and $y \simeq 0.6$ \citep[e.g.][]{StarTemps}. In either regime, we note that $M_*^{-0.25} R_*^{-1.5} \propto L_*^{-(1+6y)/4x} \approxprop L_*^{-0.35}$. So for stars from 0.4 to around 20~M$_{\odot}$, Eq.~(\ref{metfund}) reduces to

\begin{eqnarray}
{\cal D} &\approxprop& L_*^{0.4} \rp^{(3-n)} a^{-0.25} d^{-1} 
\\
&\approxprop& F_*^{0.4} \rp^{(3-n)} a^{-0.25} d^{-0.2} .
\label{basicscl}
\end{eqnarray}

An additional, temperature-dependent factor would be introduced into Eq. (\ref{basicscl}) if we were to take account of the specific wavelength of observation. In this case the flux would likely not trace the star's bolometric flux, particularly for bluer wavebands. For example, in $R$-band observations, hosts with $T \lesssim 5000$ K would be ranked lower by our metric than predicted from the approximate scaling above, as the $R$-band traces the Wien tail of their spectral energy distribution. 

It is important to stress that whilst the scaling laws above somewhat over-simplify dependencies, this is only to illustrate the broad dependencies of our decision metric. The metric itself explicitly includes the wavelength-dependent host magnitude and telescope sensitivity and therefore takes full account of the wavelength dependencies that are absent in the above scaling laws. 

Applying the same analysis for ${\cal D}_{\rm multi}$ defined by Eq.~(\ref{multi}), and using Kepler's Third Law, we recover

\be
{\cal D}_{\rm multi} \approxprop F_*^{0.4} \rp^{(3-n)} a^{-1} d^{-0.2} . \label{multiscl}
\ee

The sensitivity of the metric to, for example, habitable zone planets can be roughly gauged by considering Eq.~(\ref{metscl}) for fixed $T_{\rm eq}$:

\begin{eqnarray}
{\cal D}_{\rm habit} &\approxprop& {\cal D} \quad\quad\quad (\mbox{fixed } T_{\rm eq}) ,
\\
&\approxprop&  F_*^{0.15} \rp^{(3-n)} a^{0.25} d^{-0.7}.  \label{methab} 
\end{eqnarray}

It is perhaps unsurprising that Eqs.~(\ref{basicscl}) and (\ref{multiscl}) show that ${\cal D}$ scales almost with the square root of $F_*$, as the metric is based on signal-to-noise. ${\cal D}$ is only weakly dependent on $a$, from the product of $t_{14}^{1/2} T_{\rm eq}$. The dependence of ${\cal D}_{\rm multi}$ on $a$ is stronger here, but still potentially allows the metric to favourably rank suitable exoplanets over a useful range of host separations. We have in any case seen in Section~\ref{Extend} that the use of ${\cal D}_{\rm multi}$ over ${\cal D}$ is not necessary for determining high ranking targets. For habitable zone systems we see that ${\cal D}_{\rm habit}$ also depends on $a$ and $d$, though the additional dependencies in this form are not strong.

The scaling of Eqs.~(\ref{basicscl}) and (\ref{multiscl}) with $\rp$ clearly depends upon $n$. The most sensitive regime would be expected for Jupiter-sized planets, where we adopt $n \simeq 0$. Here the dependence is $\rp^{3}$, though a factor $\rp^2$ simply stems from the dependence of the transit method itself. For smaller planets, in the Earth- to Jupiter-size range, we have adopted $n \simeq 2$, to yield ${\cal D} \propto \rp$. This is a more modest dependence and indicates that, as transmission spectroscopy sensitivity increases to include more sub-Jupiter systems, it should become possible to detect atmospheres over a wider range of planet sizes. However, at very low planet masses we would expect the atmospheric density would become too low for detection, due to decreased surface gravity. Our decision metric assumes that the atmosphere is opaque at the observation wavelength and does not factor in the decrease in opacity that would be expected as surface gravity drops.
 
\section{Applications and Performance} 
\label{Apply}

\subsection{Single-Target Instrument Evaluation}
\label{OneTarget}

The decision metrics in Eqs.~(\ref{OldMetric}) and (\ref{multi}) are useful in several ways. Firstly, given a choice of instruments, they can be used to evaluate an optimal telescope--filter combination for a given target host star aimed at ensuring maximal ${\cal S}/{\cal N}$ for a given observing effort. The metric does not take account of the observability of individual targets from an instrument's location, or the availability of nearby reference stars. Such an ``observing cut" is carried out as part of the wider operation of the software pipeline built around our metric. 
 
\begin{table*}
	\centering
	\caption{Key information for the five telescope/instrument combinations used in this study. $\Theta_{\rm PSF}$ is the point spread function size in arcseconds. All exposure times are calculated for 50\% well-depth of one pixel for a test planet, in this case the super-Neptune WASP-127b.}
	\resizebox{\textwidth}{!}{\begin{tabular}{rlrrrrrrrrr} 
		\hline
		Telescope Name & Filter & Spectral Bins & $\lambda_{\rm cent}$ & Wavelengths & Bandwidth & Gain & m$_{\rm zp}$ & $\Theta_{\rm PSF}$ & t$_{\rm exp}$ & t$_{\rm over}$ \\ 
		 & & & (\AA{}) & (\AA{}) & (\AA{}) &(e$^-$/ADU) & & ('') & (s) & (s)\\
		\hline
		0.6m imager & Cousins $R$ & 1 & 6407 & 5500 - 7950 & 2450 & 1.15 & 22.33 & 3 & 1.4 & $\approx$ 20 \\ 
		2.4m fast imager & Sloan $r^\prime$ & 1 & 6122 & 5415 - 6989 & 1574 & 0.8 & 25.25 & 1.8 & 0.03 & $\approx$ 0\\ 
		1.5m defocused imager & Sloan $r^\prime$ & 1 & 6122 & 5415 - 6989 & 1574 & 1 & 24.57 & 14 & 37.6 & 0.58 \\
		2m spectrograph & Red arm & 20 & 7000 & 4000 - 8000 & 200 & 2.45 & 17.7 & 2.2 & 16.5 & 10\\ 
		8m spectrograph & 600RI+19 & 17 & 6777 & 5120 - 8450 & 200 & 1.25 & 27.68 & 1.38 & 8.3 & 31\\ 
		& (+GG435) & & & & & & & & &\\
		\hline
	\end{tabular}}	
	\label{InstTable}
	\label{TelConfigs}
\end{table*} 

\begin{table*}
	\centering
	\caption{${\cal D}$ scores and table positions of five example planets for each of the five test instruments, using Eq. (\ref{OldMetric}) that does not include planet mass information. Entries with no scores indicate the the set-up cannot access that target. The wider sample consists of 1558 planets, which is truncated to 552 in the case of the 2.4m fast imager}.
	\resizebox{\textwidth}{!}{\begin{tabular}{rlrrrrrrrrrr}
		\hline
		\multicolumn{2}{c}{} &
		\multicolumn{2}{c}{WASP-127b} &
		\multicolumn{2}{c}{WASP-107b} &
		\multicolumn{2}{c}{WASP-33b} &
		\multicolumn{2}{c}{HD 189733b} &
		\multicolumn{2}{c}{NGTS-1b}\\
		Telescope Name & Filter & ${\cal D}$ & \#  & ${\cal D}$ & \#  & ${\cal D}$ & \# & ${\cal D}$ & \# & ${\cal D}$ & \#\\ 
		\hline
		0.6m imager & Cousins $R$ & 2.99$\times10^5$ & 82 & 1.90$\times10^4$ & 207 & 7.48$\times10^5$ & 5 & 3.32$\times10^5$ & 62 & 5.75$\times10^4$ & 396\\ 
		2.4m fast imager & Sloan $r^\prime$ & - & - & - & - & - & - & - & - & 1.89$\times10^5$ & 20\\ 
		1.5m defocused imager & Sloan $r^\prime$ & 2.74$\times10^6$ & 40 & 1.10$\times10^6$ & 171 & 1.51$\times10^7$ & 3 & 8.98$\times10^6$ & 5 & 1.38$\times10^5$ & 471 \\
		2m spectrograph & Red arm & 2.26$\times10^4$ & 38 & 1.05$\times10^4$ & 161 & 7.73$\times10^4$ & 1 & 3.67$\times10^4$ & 17 & 1.60$\times10^3$ & 446 \\ 
		8m spectrograph & 600RI+19 & 1.55$\times10^6$ & 56 & 9.13$\times10^5$ & 175 & 4.11$\times10^6$ & 2 & 1.85$\times10^6$ & 35 & 2.10$\times10^5$ & 406\\ 
		& (+GG435) & & & & & & & & & &\\
		\hline
	\end{tabular}}
	\label{RankTable_NoMass}
\end{table*} 
	
	\begin{table*}
		\centering
		\caption{${\cal D}$ scores and table positions of five example planets for each of the five test instruments, using Eq. (\ref{MetMass}) that incorporates the planet mass.}
		\resizebox{\textwidth}{!}{\begin{tabular}{rlrrrrrrrrrr}
			\hline
			\multicolumn{2}{c}{} &
			\multicolumn{2}{c}{WASP-127b} &
			\multicolumn{2}{c}{WASP-107b} &
			\multicolumn{2}{c}{WASP-33b} &
			\multicolumn{2}{c}{HD 189733b} &
			\multicolumn{2}{c}{NGTS-1b}\\
			Telescope Name & Filter & ${\cal D}_{\rm mass}$ & \#  & ${\cal D}_{\rm mass}$ & \#  & ${\cal D}_{\rm mass}$ & \# & ${\cal D}_{\rm mass}$ & \# & ${\cal D}_{\rm mass}$ & \#\\ 
			\hline
			0.6m imager & Cousins $R$ & 1.33$\times10^6$ & 1 & 1.28$\times10^6$ & 3 & 2.77$\times10^5$ & 118 & 2.31$\times10^5$ & 154 & 5.67$\times10^4$ & 384\\ 
			2.4m fast imager & Sloan $r^\prime$ & - & - & - & - & - & - & - & - & 1.86$\times10^5$ & 26\\ 
			1.5m defocused imager & Sloan $r^\prime$ & 1.22$\times10^7$ & 3 & 7.36$\times10^6$ & 7 & 5.58$\times10^6$ & 13 & 6.25$\times10^6$ & 10 & 1.36$\times10^5$ & 471\\
			2m spectrograph & Red arm & 1.00$\times10^5$ & 1 & 7.06$\times10^4$ & 3 & 2.86$\times10^4$ & 31 & 2.55$\times10^4$ & 41 & 1.57$\times10^3$ & 439 \\ 
			8m spectrograph & 600RI+19 & 6.88$\times10^6$ & 1 & 6.14$\times10^6$ & 2 & 1.52$\times10^6$ & 82 & 1.28$\times10^6$ & 107 & 2.07$\times10^5$ & 400\\ 
			& (+GG435) & & & & & & & & & &\\
			\hline
		\end{tabular}}
		\label{RankTable_Mass}
	\end{table*}
			
Table \ref{InstTable} contains key technical information about five telescope/instrument/filter combinations used to evaluate the decision metric. They are used to define general ``classes" of telescope. These are described briefly in turn:

\begin{enumerate}
	\item PROMPT-8 --- a 0.61m robotic instrument at the Cerro Tololo Inter-American Observatory (CTIO)\footnote{\url{http://www.narit.or.th/en/index.php/facilities/southern-hemisphere-observatory/telescope}}, Chile, with a Johnson--Cousins filter set (\textit{BVRI}). PROMPT-8 is part of the SkyNet network to observe gamma-ray bursts, but has also previously been used to successfully study Rayleigh scattering in the atmosphere of the hot Neptune GJ 3470b (\citealt{GJ3470}). This telescope is used as a model for our generic \textit{0.6m imager}, representing small telescope photometry.
	\item Thai National Telescope (TNT) --- a 2.4m Ritchey-Chr\'{e}tien instrument at the Thai National Observatory (TNO) at Doi Inthanon\footnote{\url{http://www.narit.or.th/en/index.php/facilities/thai-national-observatory-tno/2-4-m-telescope/specification-tno}}, Chiang Mai, Thailand. For transit photometry, the ULTRASPEC fast-readout camera is employed in conjunction with a Sloan filter set (\citealt{ULTRASPEC}). In order to achieve negligible read-out times, the CCD must be windowed, resulting in a minimum exposure time of 2 s in order to keep a usable window on the detector. The TNT is the prototype for our \textit{2.4m fast imager} example.
	\item MuSCAT2 --- this is a new instrument installed on the 1.52m Carlos S\'{a}nchez telescope, sited at the Teide Observatory. It allows for simultaneous photometry to be undertaken in Sloan \textit{griz} filters, meaning the transmission spectrum can be studied using a single transit.  This set-up is typically operated with strong defocusing (\citealt{MuSCAT2}), and forms the basis for our \textit{1.5m defocused imager}.
	\item Liverpool Telescope (LT) --- a 2m-aperture robotic instrument at the Observatorio del Roque de Los Muchachos (ORM), La Palma. The configuration we have chosen here is the Spectrograph for the Rapid Acquisition of Transients (SPRAT)\footnote{\url{http://telescope.livjm.ac.uk/TelInst/Inst/SPRAT/}}, employing the red-optimised arm of the instrument. This set-up represents our \textit{2m spectrograph}.
	\item Very Large Telescope (VLT) --- employing an 8m-class instrument for broad-band photometry is generally not useful, except for particularly faint hosts or small planets. Better precision can instead be achieved with a spectrograph, so the FOcal Reducer/low dispersion Spectrograph 2 (FORS2) suite\footnote{\url{http://www.eso.org/sci/facilities/paranal/instruments/fors/overview.html}} is employed in our comparison here, and forms the basis for our \textit{8m spectrograph}.
\end{enumerate}

In order to test the capabilities of our five instruments against each other, the metric was run using Eq. (\ref{OldMetric}) for a test planet. The chosen target was the bloated super-Neptune WASP-127b (\citealt{WASP-127Disc}), a low-density world in orbit around a moderately bright G5 star ($V$ $\approx$ 10.15), which is now thought to have a relatively cloudless atmosphere (\citealt{WASP-127bSpectro}, \citealt{WASP-127bSpectro2}). In these respects, it is an ideal target for transmission spectroscopy. 

For the purposes of comparative analysis, Cousins $R$ and Sloan $r'$ filters are employed for all of our set-ups; failing that, filters/grisms that match these as closely as possible are chosen. In practice, it is of course necessary to observe transits using several filters/bands, ideally in both the optical and near infra-red regimes, in order to construct an atmospheric spectrum through photometry. For focussed observations we adopt the average seeing from the site, assuming a Gaussian point spread function. However, the 1.5m imager uses a defocus such that $\Theta_{\rm PSF} = 14''$ (\citealt{MuSCAT2}). We use a top-hat distribution to model this. For spectrographs, the dispersed light is treated as a series of elements 1 pixel across, which are then modelled as Gaussian in intensity, with a width dependent on $\Theta_{\rm PSF}$.

The two spectrographic set-ups both use slits of $10''$ width and are set to a common bin width of 200\AA{}. The 1.5m defocused imager and 2m spectrograph use average per-channel $m_{\rm zp}$ values; all other cases employ broad-band zero points (converted appropriately in the case of the 8m spectrograph).

The exposure time for each instrument was chosen such that an individual CCD well would be 50\% filled, thus facilitating comparison between the different telescopes. These times are collected together in Table \ref{InstTable}, while Table \ref{RankTable_NoMass} holds the scores for some example targets. Our predicted exposure times were verified against available exposure time calculators\footnote{\url{http://telescope.livjm.ac.uk/TelInst/calc/}}\textsuperscript{,}\footnote{\url{http://www.eso.org/observing/etc/bin/gen/form?INS.NAME=FORS+INS.MODE=spectro}}, existing data frames and through communications with instrument teams (Narita, private communication).

Inspecting Table 2, the 2.4m fast imager is too sensitive to target WASP-127 while operated in a focused mode and is restricted to fainter targets. Our defocused imager returns the highest ${\cal D}$ score, while the 2m spectrograph returns the lowest. These two instruments are of comparable aperture, but the spectrograph's lower ${\cal D}$ scores reflect its much higher spectral resolution.

Due to the relative brightness of WASP-127, only short exposure times are needed, particularly in the cases of the 0.6m imager and 8m spectrograph, meaning we approach the regime where \textit{t}$_{\rm exp} \ll \textit{t}_{\rm over}$, as discussed in Section \ref{Extend}. For these set-ups, the target's $\cal D$ score is suppressed due to this; WASP-127b appears further down the list compared to the other two set-ups. The exercise demonstrates that a target may not necessarily be suitable for all instruments, illustrating our metric's utility in identifying the most appropriate telescope asset to employ for a given target. However, WASP-127b remains in the top 6\% of targets for four of the five telescopes tested here, making it a high priority target.

We can repeat this exercise for four other planets; the hot Jupiters WASP-33b (\citealt{WASP33Disc}), HD 189733b (\citealt{HD189733Disc}) and the recently-discovered NGTS-1b (\citealt{NGTS-1b_Disc}), in orbit around A5, K-type and M-dwarf stars respectively, and the super-Neptune WASP-107b (\citealt{WASP-107Disc}). Again, the 2.4m focused imager is unable to access the three bright targets, as expected. However, it returns the highest $\cal D$ score for NGTS-1b $(V = 15.5)$, clearly making it the preferred choice for accessing faint targets while maintaining a high cadence. Otherwise, WASP-33b and HD 189733b rank among the best targets to study, with one notable exception; the 0.6m imager ranks HD 189733b 30 places further down the list compared to the other four instruments. This is due to the brightness of the host star ($V$ = 7.7, listed in \citealt{HD189733Disc}), which means that we enter the regime of \textit{t}$_{\rm exp} \ll \textit{t}_{\rm over}$, strongly suppressing the returned $\cal D$. This can of course be countered by defocusing, as demonstrated with the high score given by the 1.5m defocused imager. WASP-33b is also in orbit around a bright host ($V$ = 8.3, listed in \citealt{WASP33Disc}), but its far higher equilibrium temperature, longer transit duration and larger radius (in comparison with HD 189733b) means it still places at the top of the table.  WASP-107b meanwhile ranks much further down the table for the remaining four set-ups, due to its fainter host of $V$ = 11.6 and predicted equilibrium temperature of 770K.

The masses of all five of these planets are known, meaning we can also employ the mass-sensitive metric ${\cal D}_{\rm mass}$ defined in Eq. (\ref{MetMass}) and make a direct comparison. The scores and table positions resulting from this are in Table \ref{RankTable_Mass}. We can see immediately from this that targets known to have bloated atmospheres (WASP-107b and -127b) now place at the top of the table, while the more massive WASP-33b is downweighted, significantly in the cases of the 0.6m imager and 8m spectrograph. Comparing WASP-33b in turn to HD 189733b, we can expect the latter to have a stronger signal due to its greater transit depth. For instruments where $t_{\rm over} \approx 0$ (fast imagers), or where exposure times can be made sufficiently long by defocusing, we see that HD 189733b is indeed predicted to be a better target. For our remaining set-ups in the regime of $t_{\rm exp} \ll t_{\rm over}$, this target is still downweighted below WASP-33b for this reason.

\subsection{Accessible Populations of Transiting Planets}
\label{CutoffMade}

While our decision metric can be used for single targets, it is far more useful to extend its application to the entire population of transiting planets, and produce a ranked table which can be used to short-list targets for a particular instrument. This can allow observers to assess the relative potential of different telescope assets for transmission spectroscopy studies.

Unfortunately, an absolute figure of merit cannot be recovered, due to the fact that our metric does not have an absolute sensitivity limit for atmospheric detections, stemming from the approximate proportionality in the \textit{H}-scaling in Eq. (\ref{hscale}). Hence, when comparing set-ups, the best we can do is to select one set-up as a reference, and then compare others to it for some subjective cut-off, referred to as $\cal D_{\rm cut}$. In this way, it can be used to gauge which regions of parameter space are preferred for a particular instrument, as well as comparing the relative performance of different set-ups.

Using the five set-ups described in Table \ref{InstTable}, the metric score $\cal D$ is calculated for all transiting planets. The initial intake of planets is provided by TEPCat\footnote{\url{http://www.astro.keele.ac.uk/jkt/tepcat/tepcat.html}}, described in \cite{TEPCatPaper}. This catalogue contains an extensive set of physical and observational properties for all confirmed exoplanets and brown dwarfs. 

\begin{figure*}
	\includegraphics[width=\columnwidth]{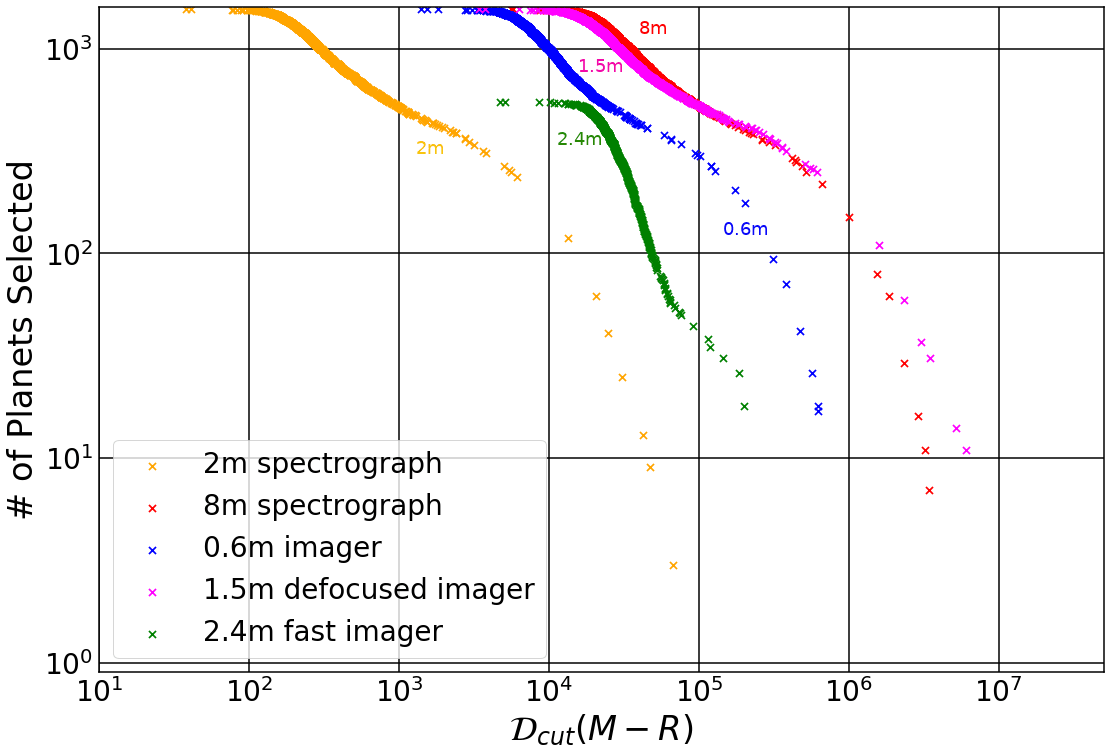}
	\includegraphics[width=\columnwidth]{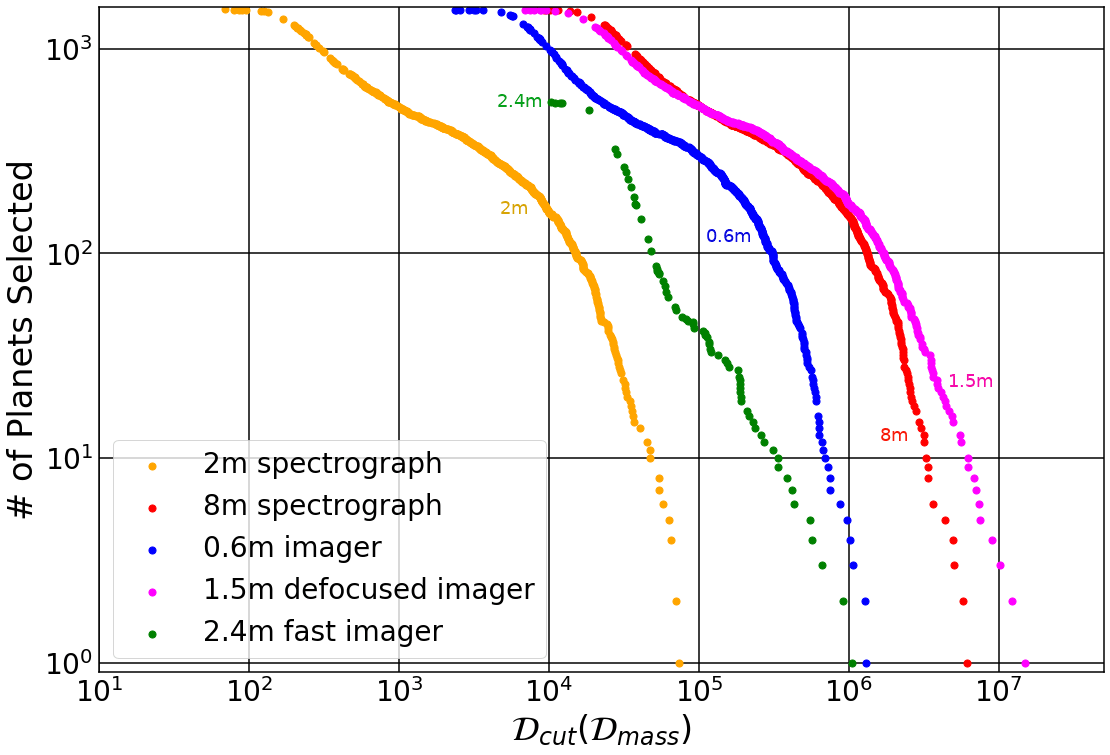}
	\caption{The dynamic range of $\cal D$ scores recovered for the telescope configurations listed in Table~\ref{TelConfigs}, illustrating the number of potential transmission spectroscopy targets for a given $\cal D_{\rm cut}$. The left panel is for all targets using the two-part mass-radius relation, while the right is for targets with known masses.}
	\label{DynamicCut}
\end{figure*}

Figure \ref{DynamicCut} shows these scores plotted for each of our five set-ups, and how the number of planets selected varies with threshold $\cal D_{\rm cut}$. To facilitate a like-for-like comparison, two calibration instruments are used; the 1.5m defocused imager for photometry (purple points) and the 8m spectrograph for spectrographic set-ups (red points). Most of our instruments obey the same general distribution, but the sample of targets for the 2.4m fast imager is truncated due to the minimum exposure time restriction outlined in the previous section.

In order to make a ``viability cut", the sample of planets for the calibration instrument is first sorted by $\cal D$ score, highest to lowest. Working down the sorted list, the $\cal D$ scores are summed until a chosen fraction of the total cumulative score is reached. All planets that comprise this chosen fraction of the total cumulative score form the ``viable sample" for the calibration instrument. This fraction is referred to as $f (\cal D > \cal D_{\rm cut})$, from which the cut-off value is recovered. This ${\cal D}_{\rm cut}$ value is then applied to the other instruments to determine their respective viable samples.

\begin{figure*}
	\includegraphics[width=\columnwidth]{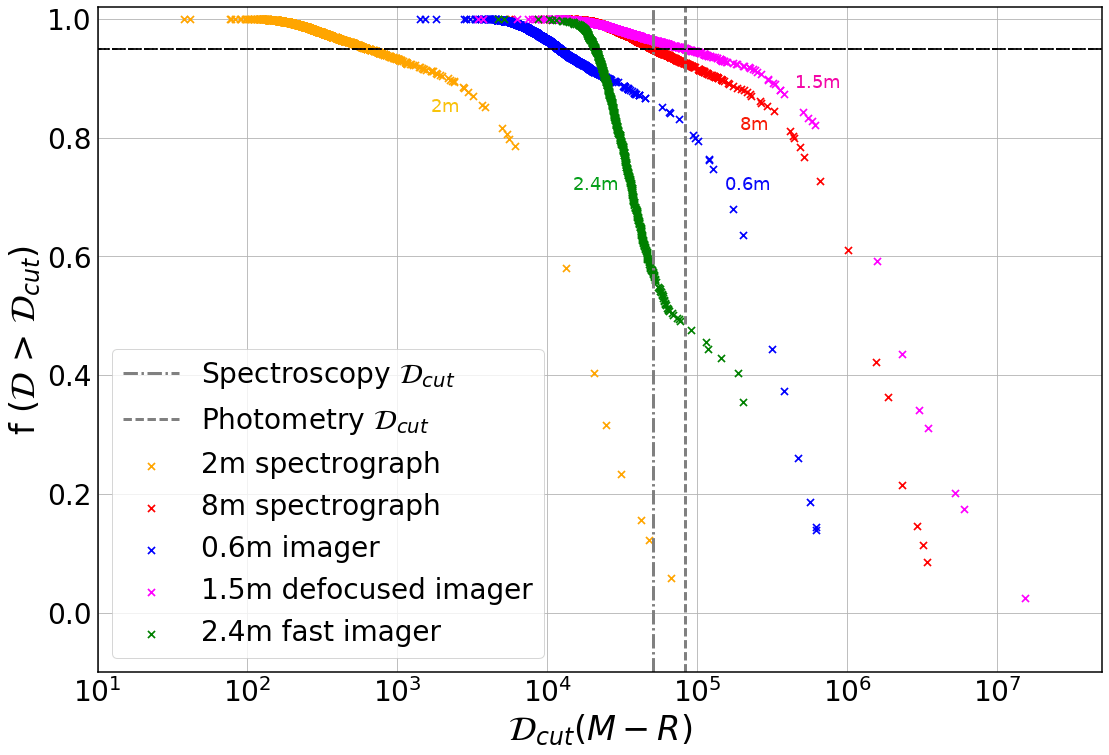}
	\includegraphics[width=\columnwidth]{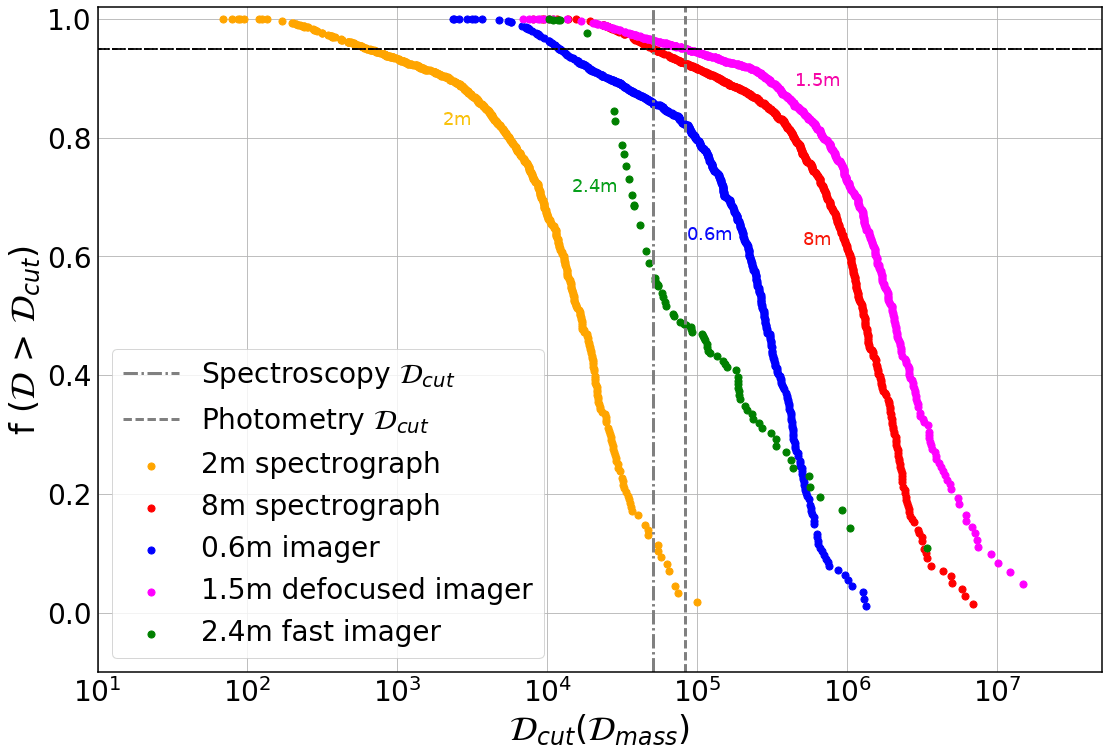}
	\caption{The cumulative fraction $f$ of planets with ${\cal D} > {\cal D}_{\rm cut}$ as a function of ${\cal D}_{\rm cut}$. Two cut-offs, one for spectroscopy and one for photometry, are indicated by the vertical lines. These are chosen to give $f = 0.95$ for the 8m spectrograph and for the 1.5m defocused imager, respectively.}
	\label{SumFracCut}
\end{figure*}

Figure \ref{SumFracCut} shows the results of this exploration. The y-axis of Figure \ref{SumFracCut} shows the fraction of the total cumulative $\cal D$ score that is above our cut-off value. It is apparent that the cumulative distribution has a distinct ``knee", beyond which is a long tail of planets with lower scores. Ultimately, the choice of $\cal D_{\rm cut}$ must be validated through observations. However, for the purposes of this exercise, $\cal D_{\rm cut}$ is chosen in order to allow us access to the relative performance of all five instruments for comparison. Although not a observationally-motivated cut (as described in Section \ref{moreNoise}), it is a sensible cut for this purpose.

To this end, $\cal D_{\rm cut}$ is computed such that $f (\cal D > \cal D_{\rm cut})$ = 0.95 for our calibration instruments, with the number of planets yielded for each set-up displayed in Figure \ref{SumFracCut}. These are the samples used in our subsequent analysis. Considering for example our photometric set-ups, the fact that the samples for the 1.5m and 0.6m imagers are not identical defines a sub-set of targets that the larger instrument can access, but that the smaller one may not be able to.

\begin{figure*}
	\includegraphics[width=\columnwidth]{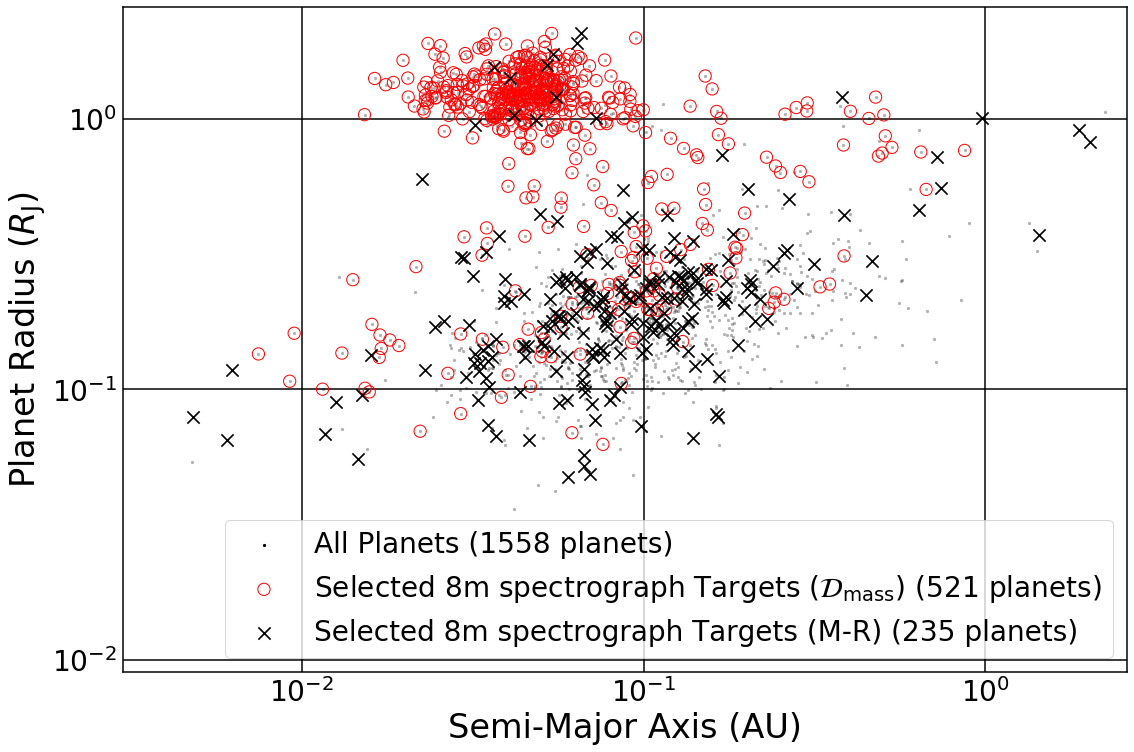}
	\includegraphics[width=\columnwidth]{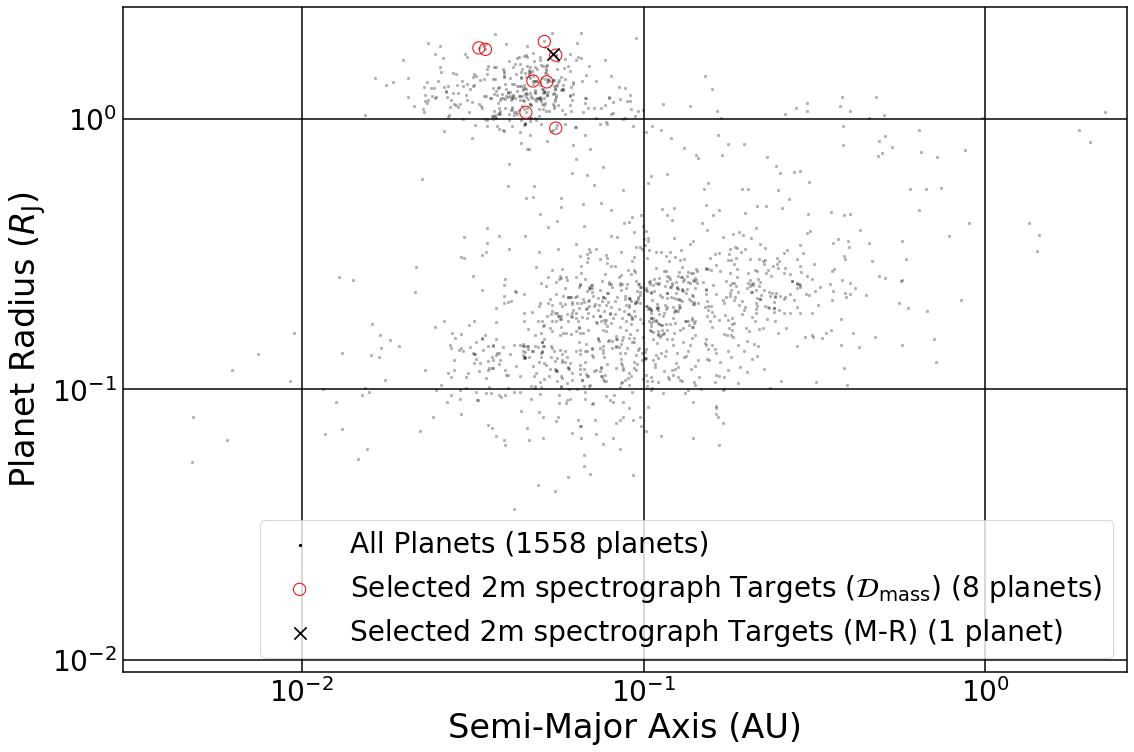}
	\includegraphics[width=\columnwidth]{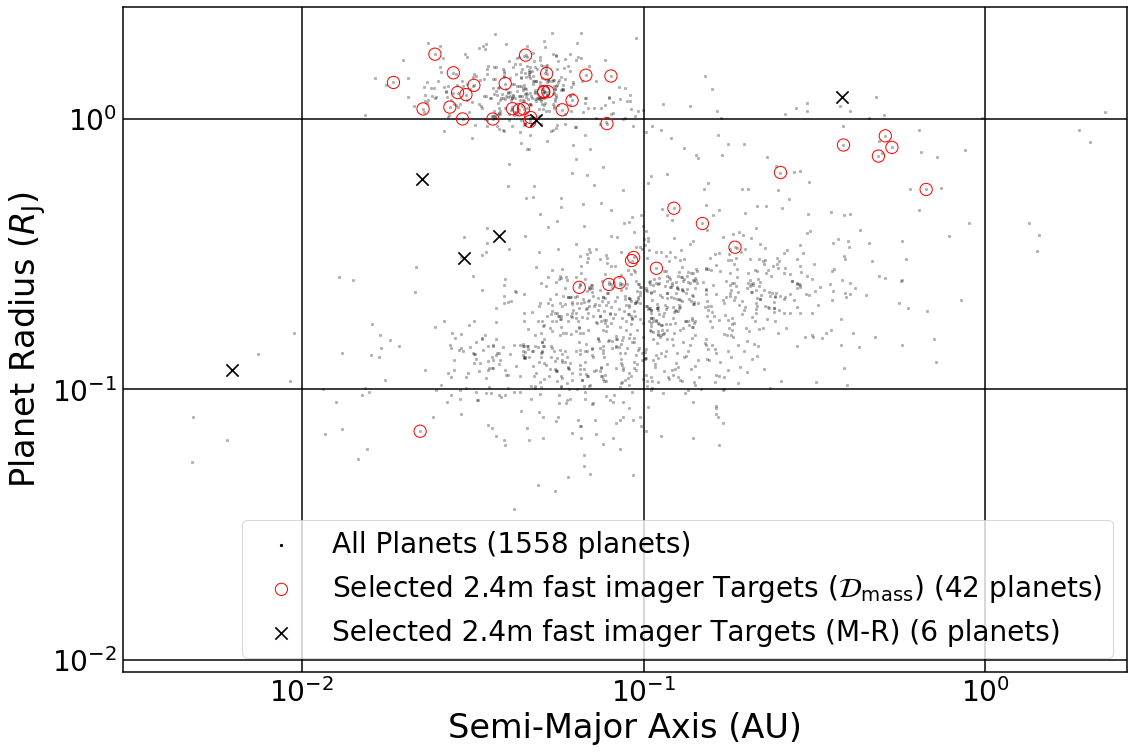}
	\includegraphics[width=\columnwidth]{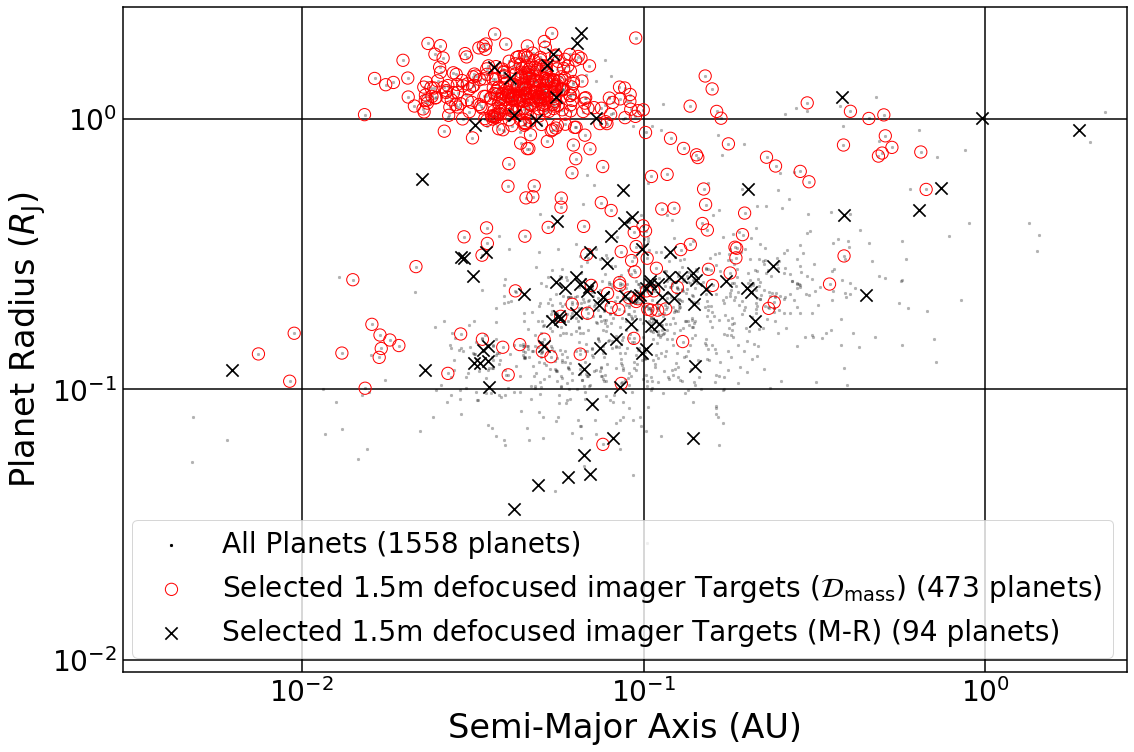}
	\includegraphics[width=\columnwidth]{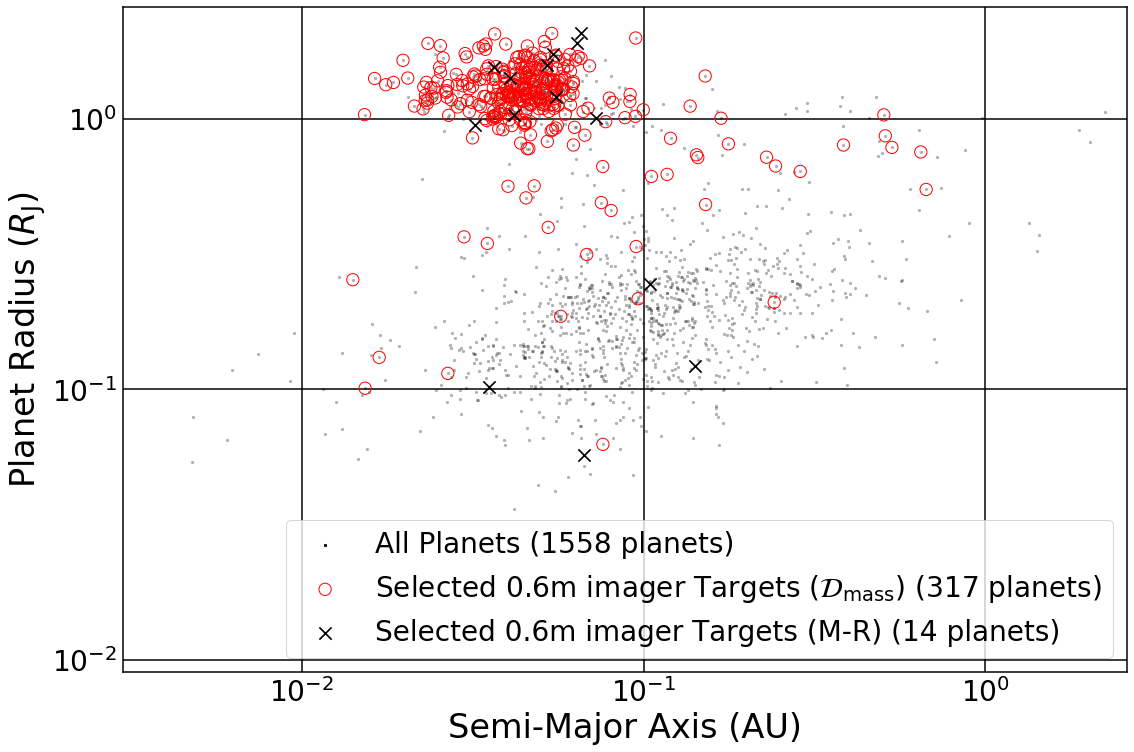}
	\caption{Planet radius against semi-major axis for selected transmission spectroscopy targets for each of the telescope configurations listed in Table~\ref{TelConfigs} (coloured markers). The selected targets are superimposed on the complete sample of 1558 planets (grey points). To show the relative effectiveness of each imager or spectrograph we have defined the selected target list as those that collectively comprise more than $95\%$ of the cumulative metric score for our reference instruments (1.5m defocused imager and 8m spectrograph, respectively -- see Figure~\ref{SumFracCut}). Red markers are selected targets for which we have mass estimates, while black markers are selected targets for which the M-R relation is employed.}
	\label{RpVa}
\end{figure*}

\begin{figure*}
	\includegraphics[width=\columnwidth]{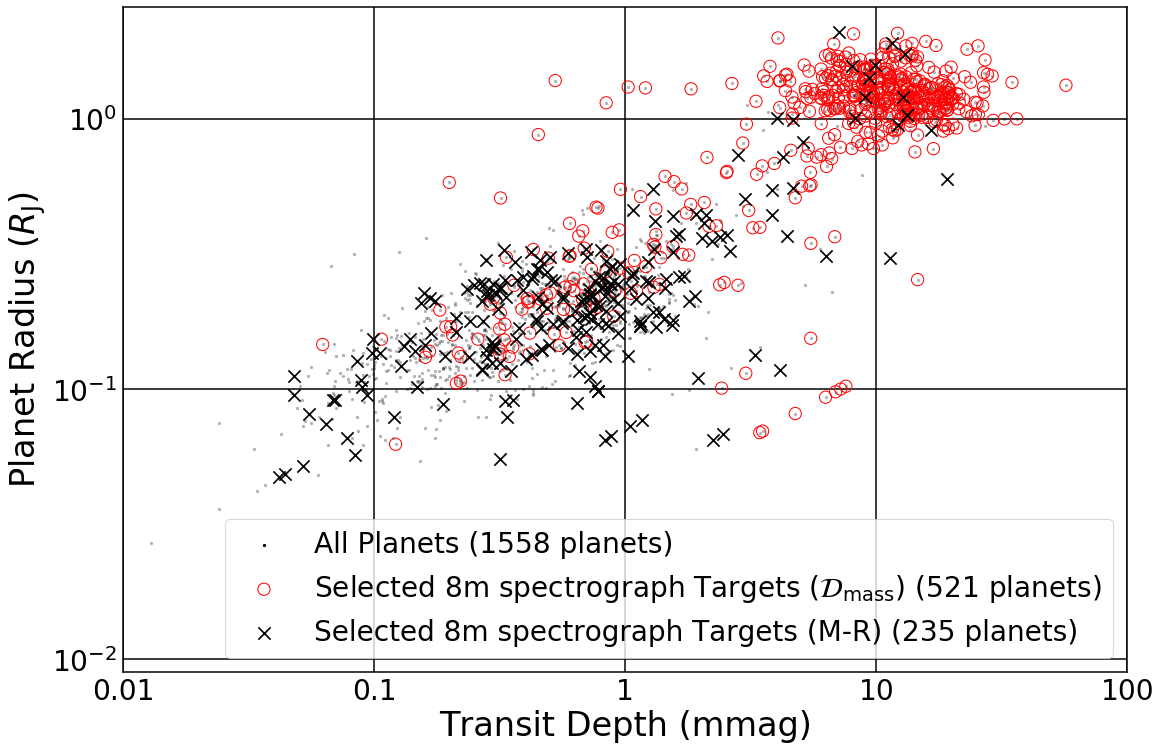}
	\includegraphics[width=\columnwidth]{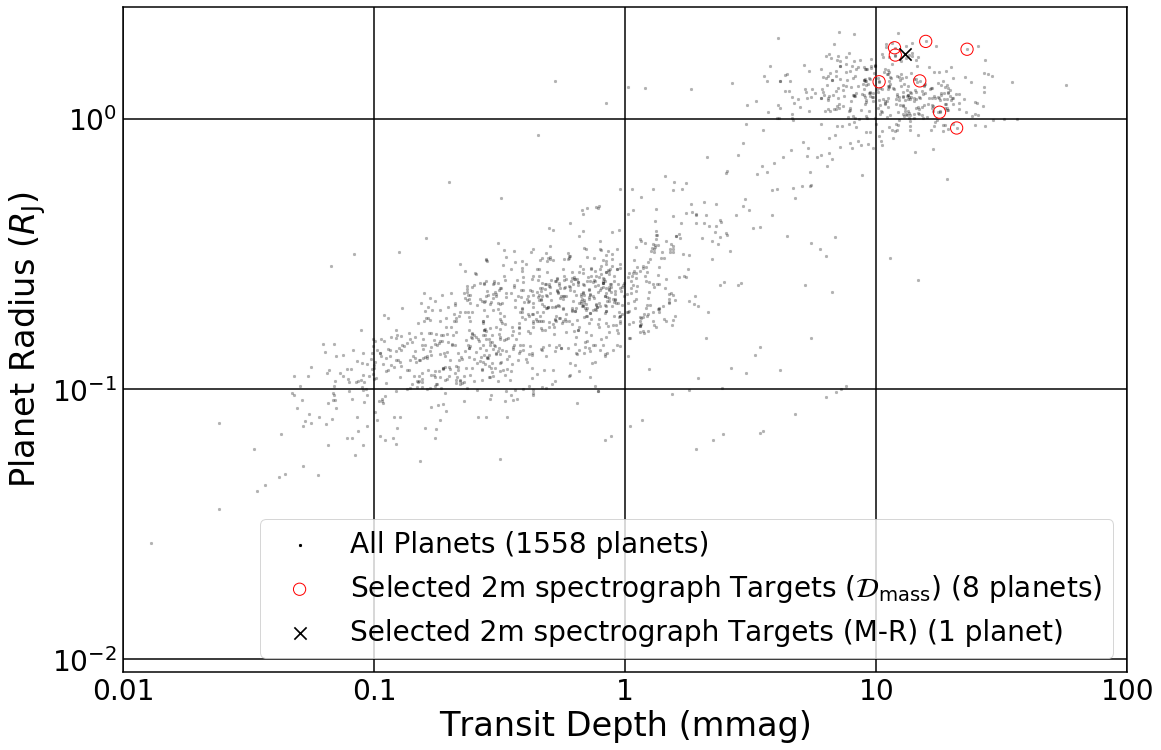}
	\includegraphics[width=\columnwidth]{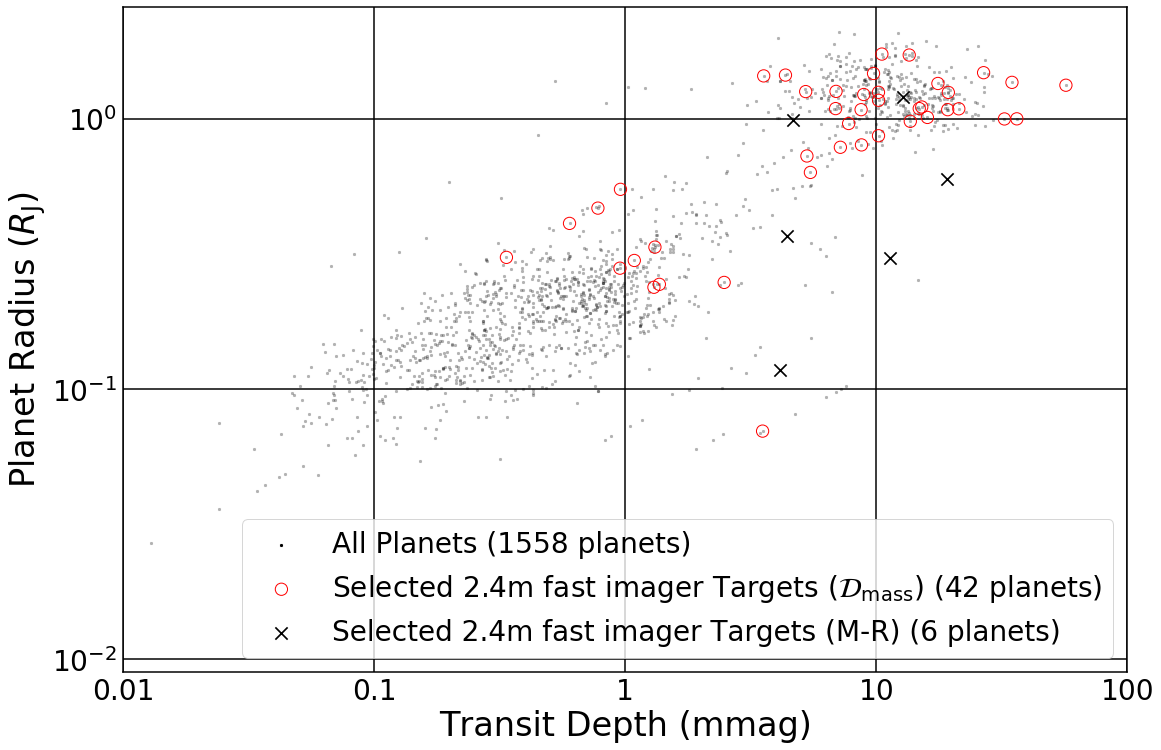}
	\includegraphics[width=\columnwidth]{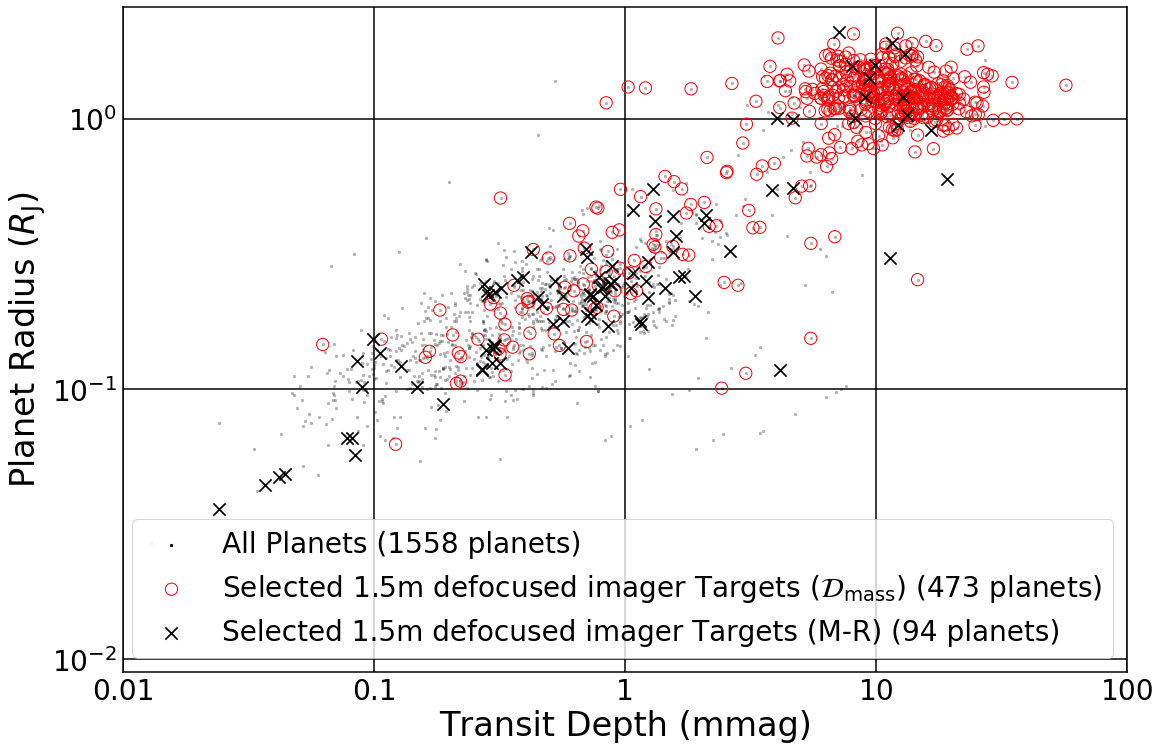}
	\includegraphics[width=\columnwidth]{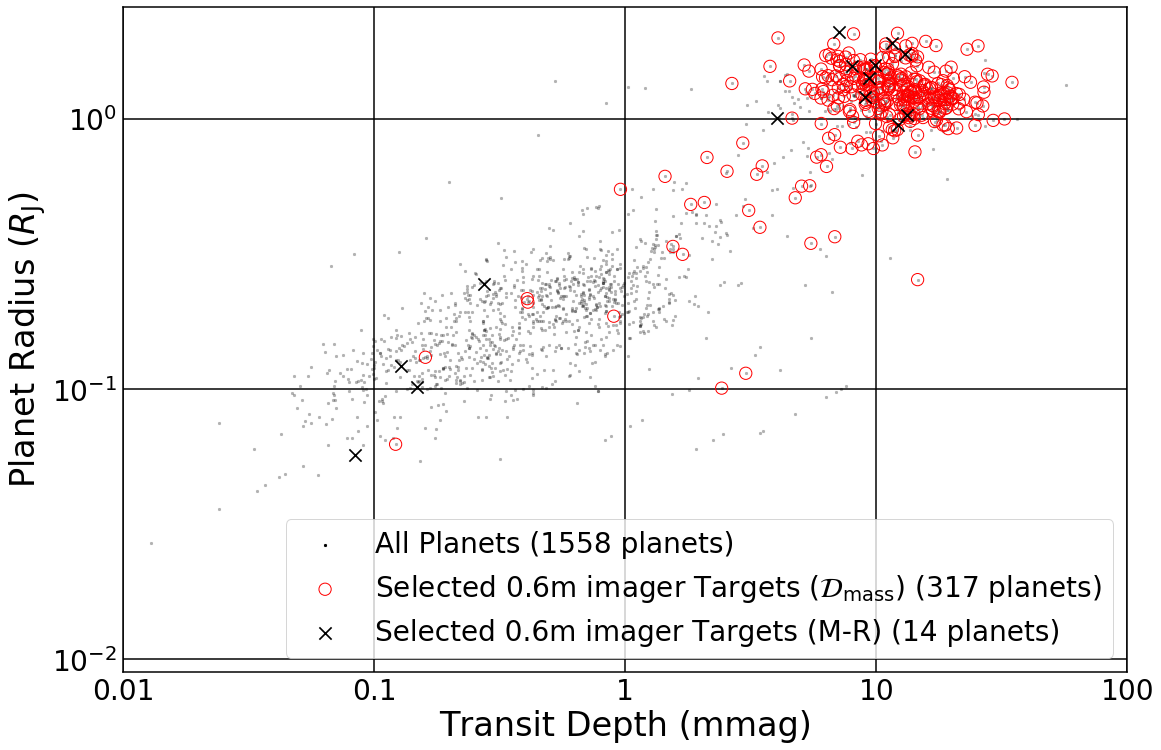}
	\caption{Planet radius against transit depth for selected transmission spectroscopy targets, with symbols as in Figure~\ref{RpVa}.}
	\label{RpVD}
\end{figure*}

With this sensitivity cut made, it is now useful to explore the size of each sample and their distribution across parameter space, in order to check that the selection is truly representative. Figures \ref{RpVa} and \ref{RpVD} shows plot of planet radius against semi-major axis and transit depth respectively for our five recovered samples. There exists a large population of planets which are not accessible to any of our set-ups; this consists largely of \textit{Kepler} planets orbiting stars too faint for effective follow-up. As such, these planets do not have known masses, making up the bulk of the population in the left panels of Figs. \ref{DynamicCut} and \ref{SumFracCut}. The hot-Jupiter-type population of planets is accessible to all five of our instruments, demonstrating that even small telescopes can be effectively used to study such targets. These typically produce deep transits and are also conducive to RV follow-up, allowing masses to be recovered and $\cal D_{\rm mass}$ to be employed.

Under our chosen $\cal D_{\rm cut}$, four of our instruments can access Jupiter-sized bodies at larger separations, but proportionally fewer longer-period objects are selected than short-period. The transit technique is inherently biased towards detecting large, close-in planets, so this can be reasonably expected. Our metric scales with \textit{a}$^{-1/4}$, but this weak dependency is not enough by itself to produce this selection effect. Objects in close-in orbits also have high $T_{\rm eq}$ values, and ${\cal D} \propto T_{\rm eq}$; it is this stronger co-dependency that causes long-period objects to be disfavoured.

\begin{figure*}
	\includegraphics[width=\columnwidth]{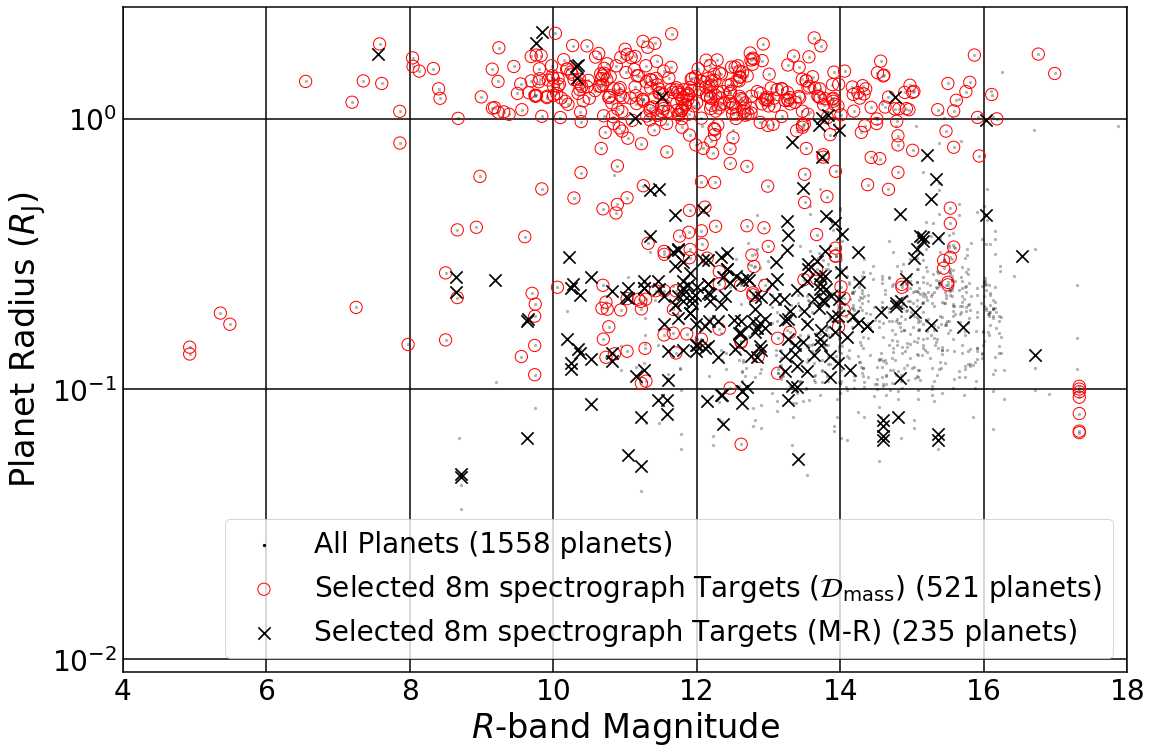}
	\includegraphics[width=\columnwidth]{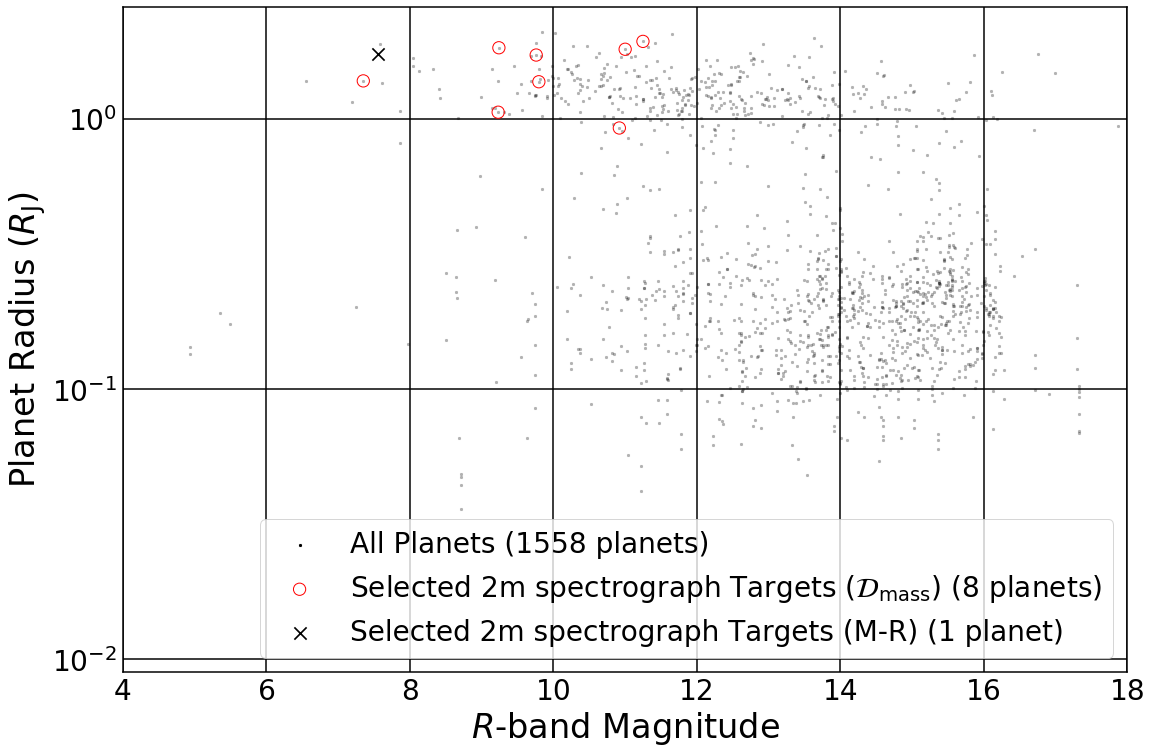}
	\includegraphics[width=\columnwidth]{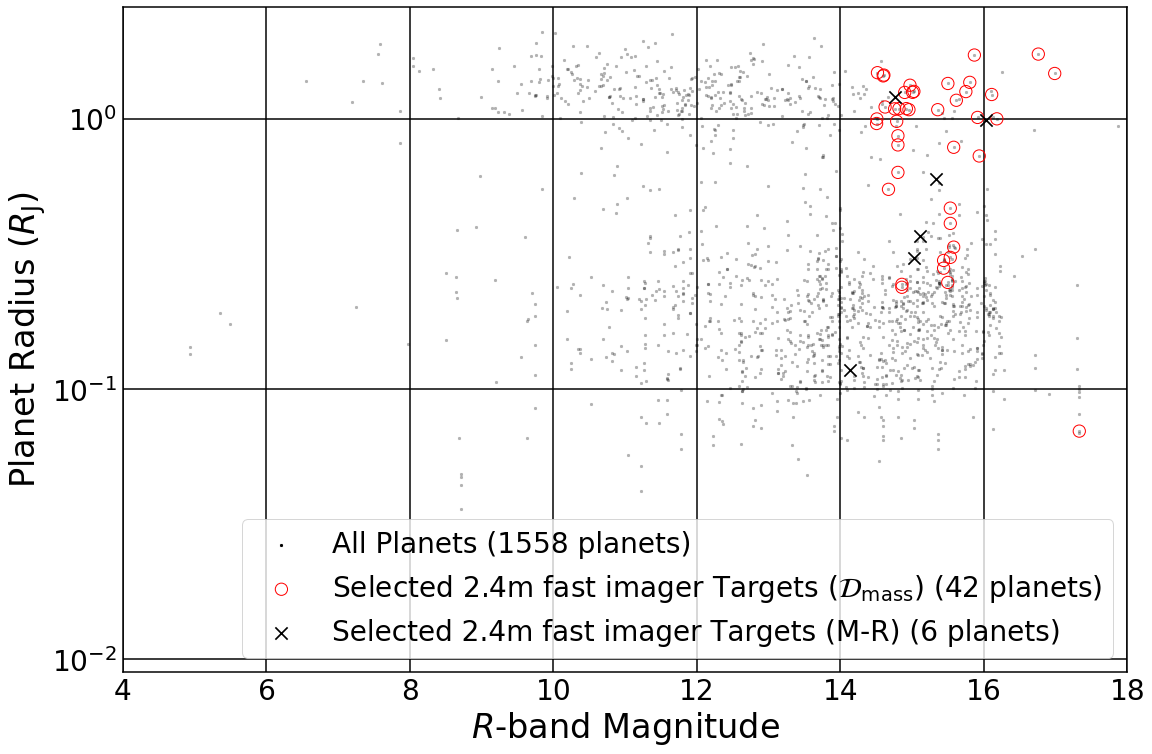}
	\includegraphics[width=\columnwidth]{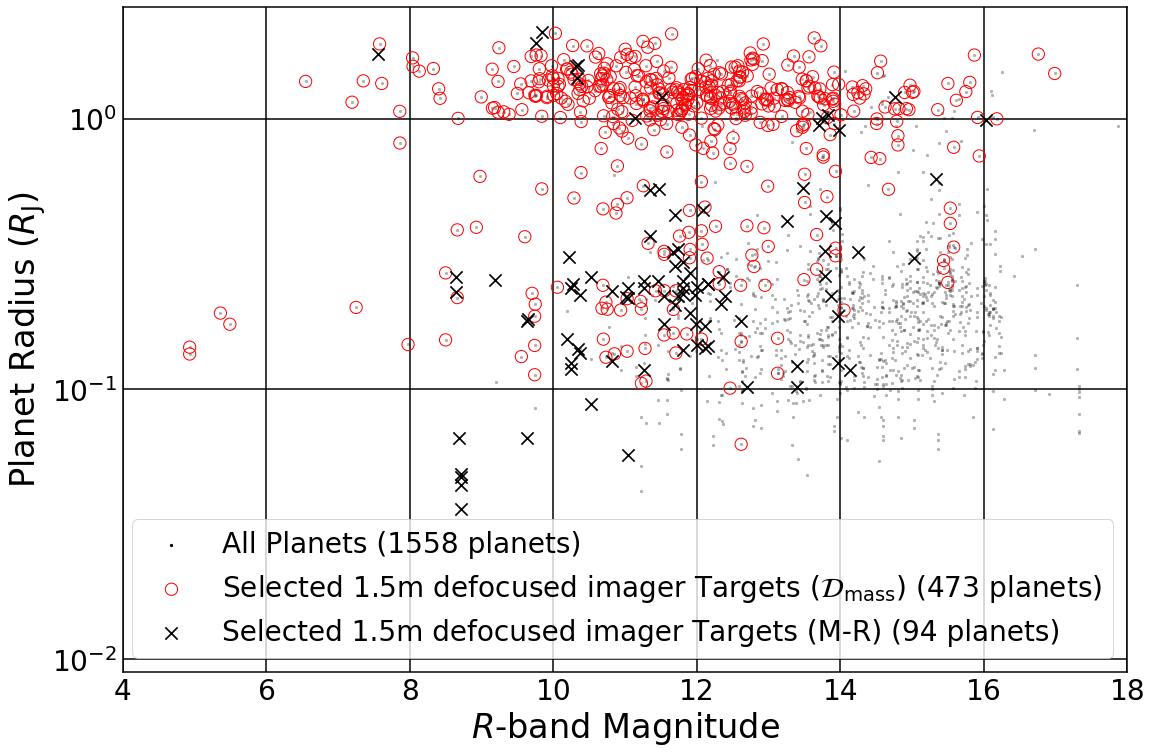}
	\includegraphics[width=\columnwidth]{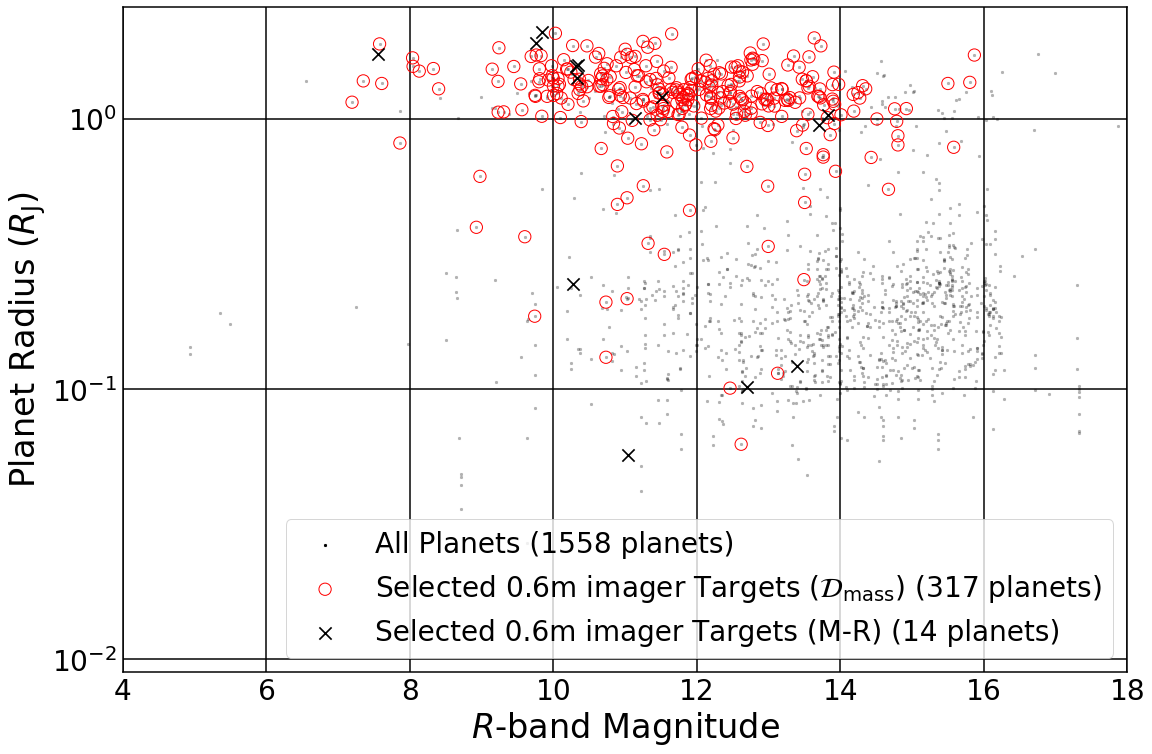}
	\caption{Planet radius against $R$-band magnitude for selected targets, with symbols as in Figure \ref{RpVa}.}
	\label{RpVRMag}
\end{figure*}

Figure~\ref{RpVRMag} shows planet radius against $R$-band magnitude. We see a stratified distribution here in apparent $R$-band magnitude; under this cut, the 2m spectrograph can access only a small set of bright targets, while the larger spectrograph and two of the imagers can include a much larger sample, spanning the distribution of parent star magnitudes. Conversely, the focused 2.4m fast imager is restricted to faint targets, in order to avoid saturation and keep above the minimum required exposure time.

\begin{figure*}
	\includegraphics[width=\columnwidth]{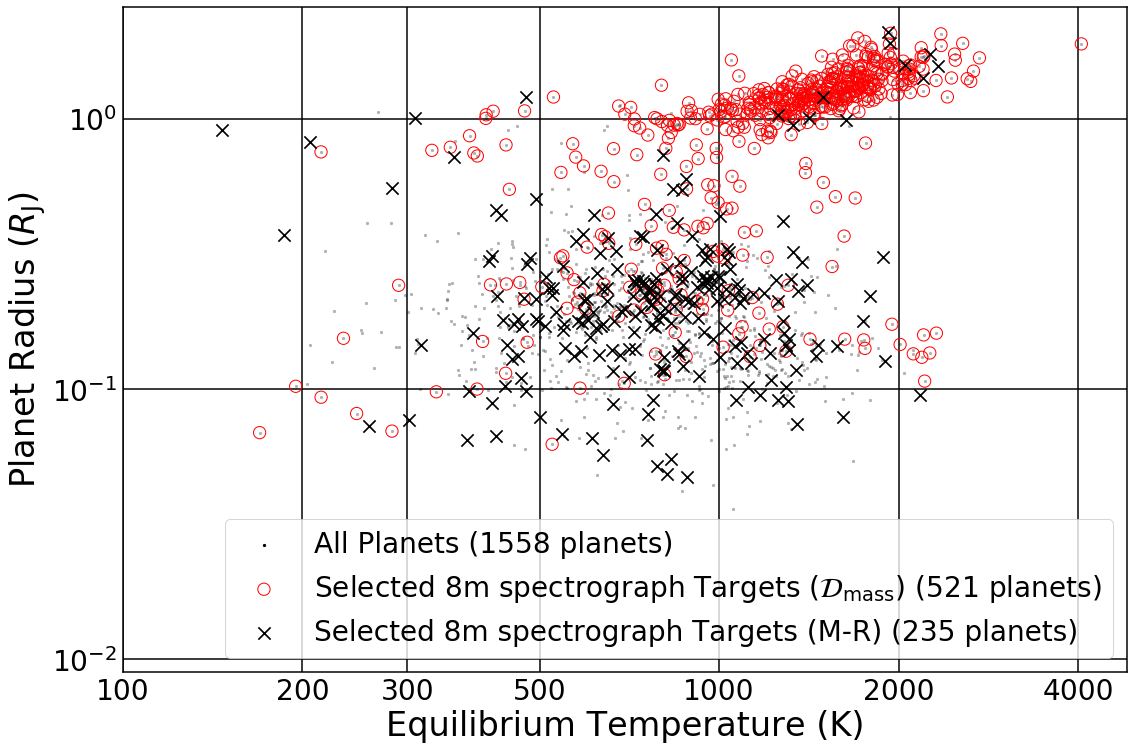}
	\includegraphics[width=\columnwidth]{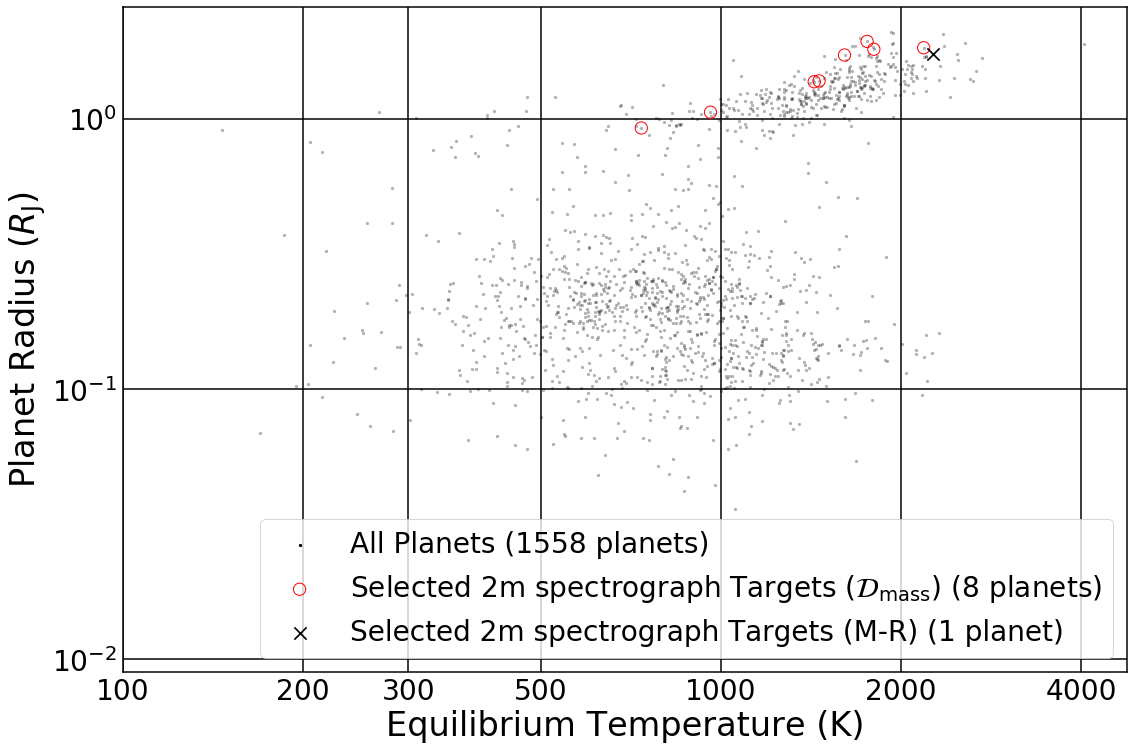}
	\includegraphics[width=\columnwidth]{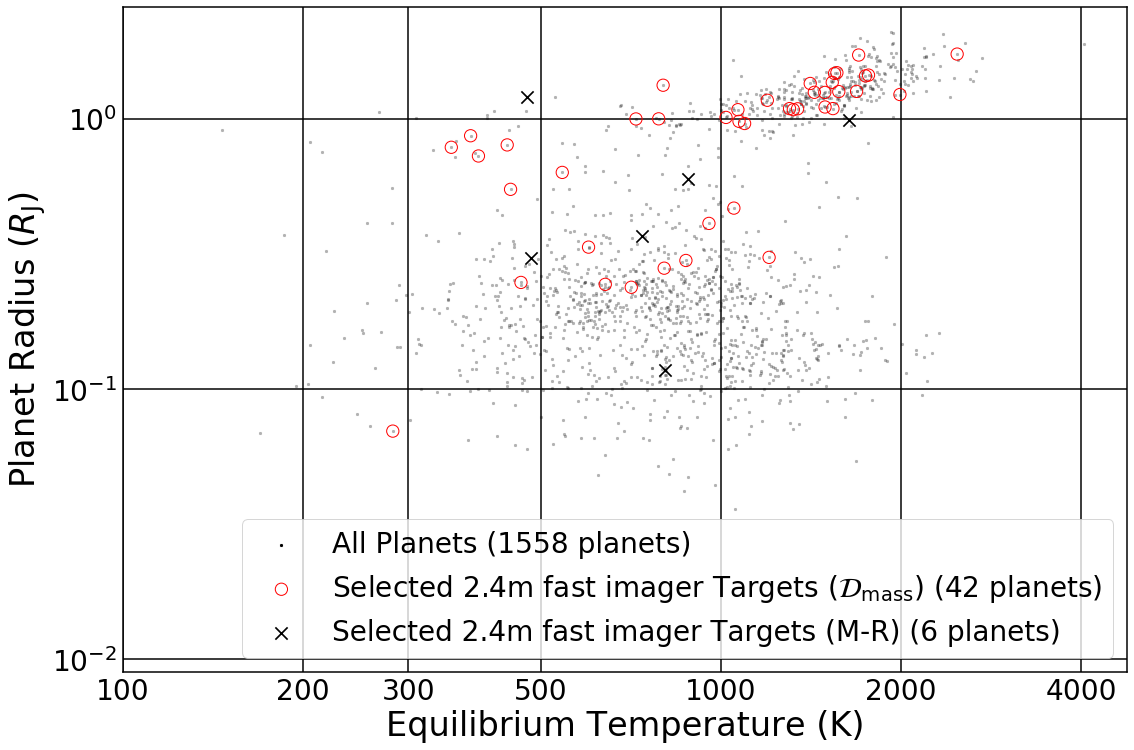}
	\includegraphics[width=\columnwidth]{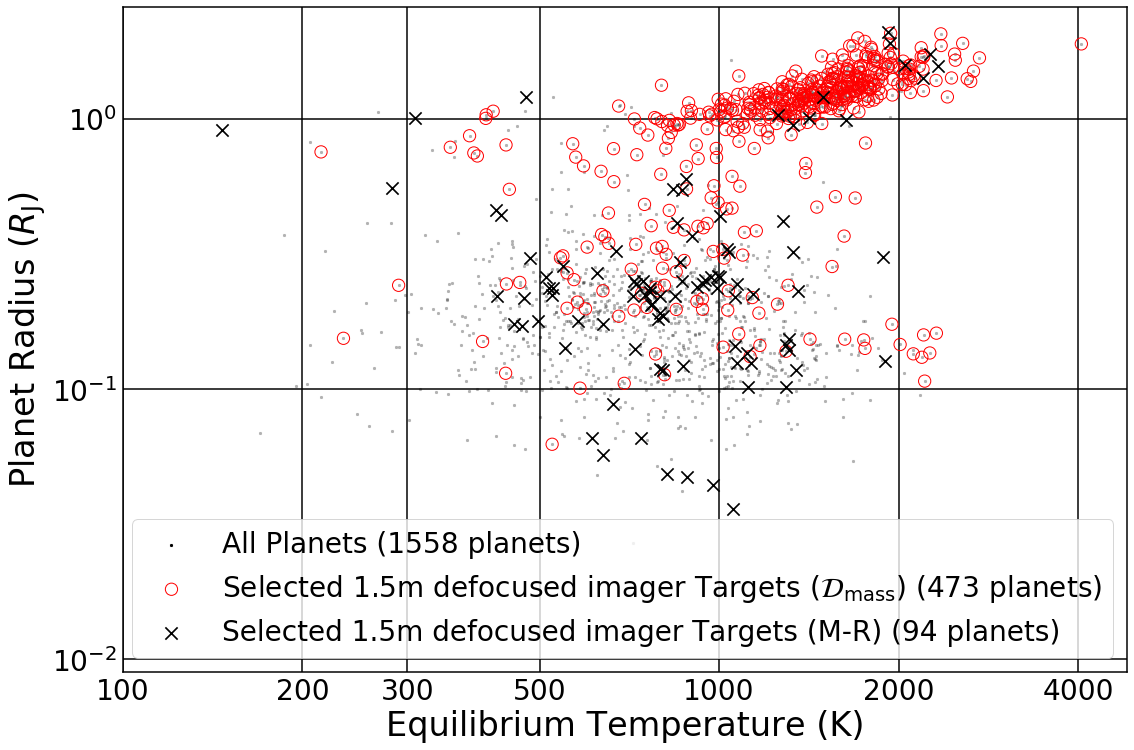}
	\includegraphics[width=\columnwidth]{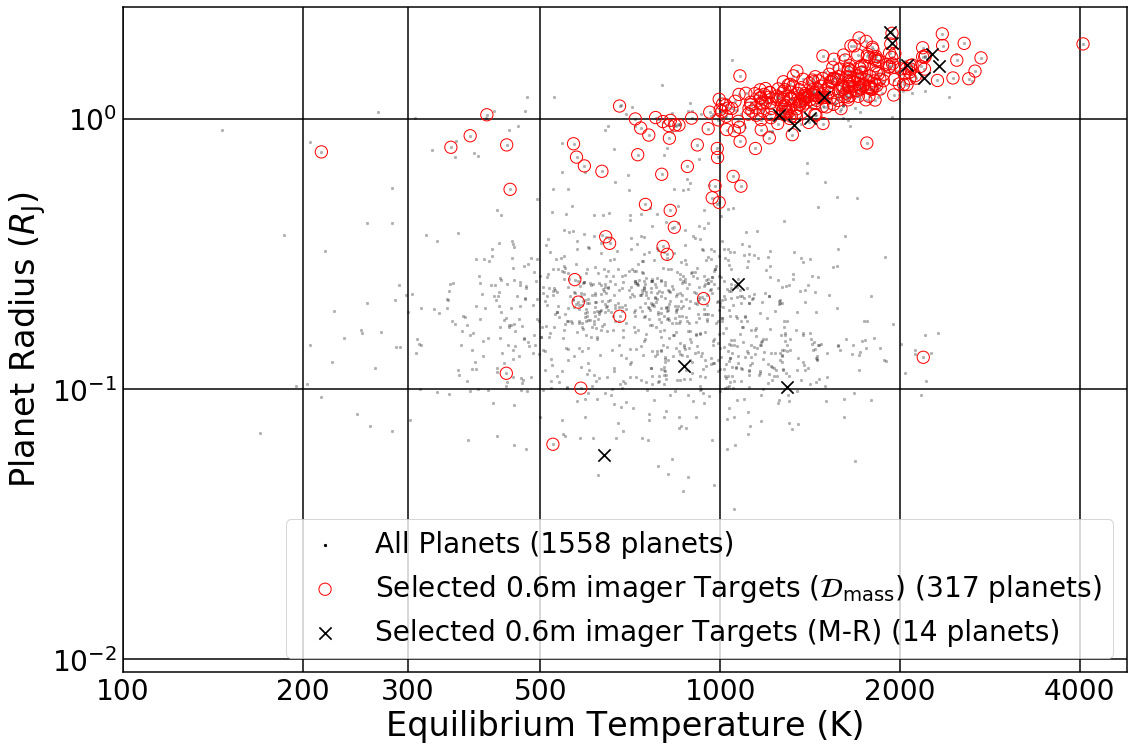}
	\caption{Planet radius against equilibrium temperature for selected targets, with symbols as in Figure \ref{RpVa}.}
	\label{RpVT}
\end{figure*}

Figure \ref{RpVT} shows our samples in the plane of planet radius against equilibrium temperature. We know from Section \ref{bias} that the metric has a linear dependency on \textit{T}$_{\rm eq}$, but our set-ups are able to sample targets over a broad temperature range. The lone planet at \textit{T}$_{\rm eq} \approx$ 4100K is the ``super-puff" KELT-9b (\citealt{KELT-9bDisc}), a very hot Jupiter in a close-in orbit around an A-type star. If other similar planets are discovered, e.g. by \textit{TESS}, it follows that even small telescopes should be able to access this new population.

\subsection{Comparison with Previous Target Selections} \label{compare}

As stated in Section \ref{Intro}, selection of promising targets for atmospheric study has been predominantly done manually, often based on their being identified as good targets for atmospheric study by the discoverers. If our metric is to be useful it should score previously studied systems highly. Efforts are underway to define target samples for future transmission spectroscopy studied with facilities such as the \textit{James Webb Space Telescope} \citep[\textit{JWST},][]{JWSTSynth}. An objective decision metric can play an important role in such selections. It is a useful exercise then to review which planets have been studied so far, or that have been selected for future study to see if they score highly and are ranked as good candidates by our decision metric. 

\begin{figure*}
	\includegraphics[width=0.77\textwidth]{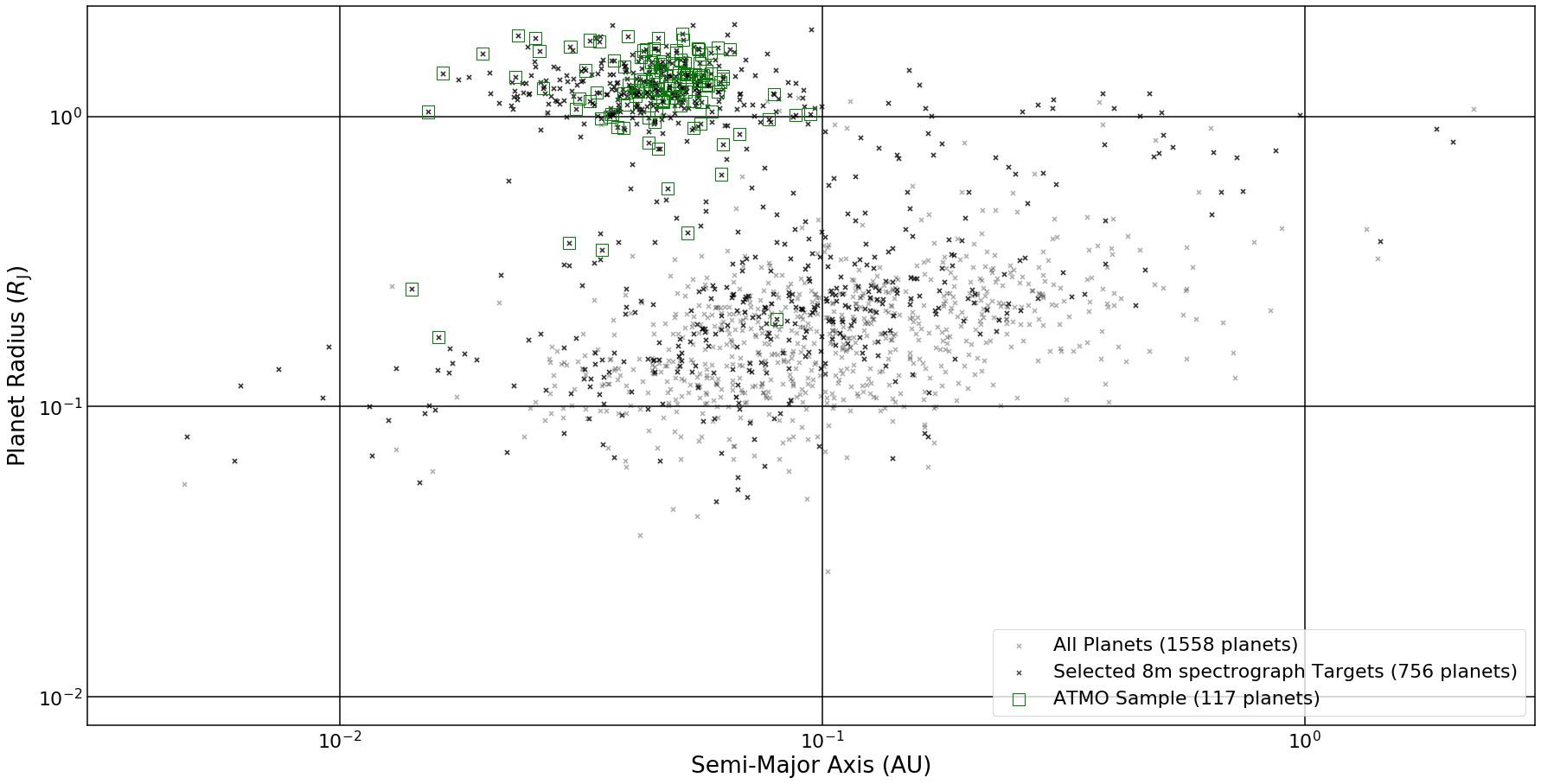}
	\includegraphics[width=0.77\textwidth]{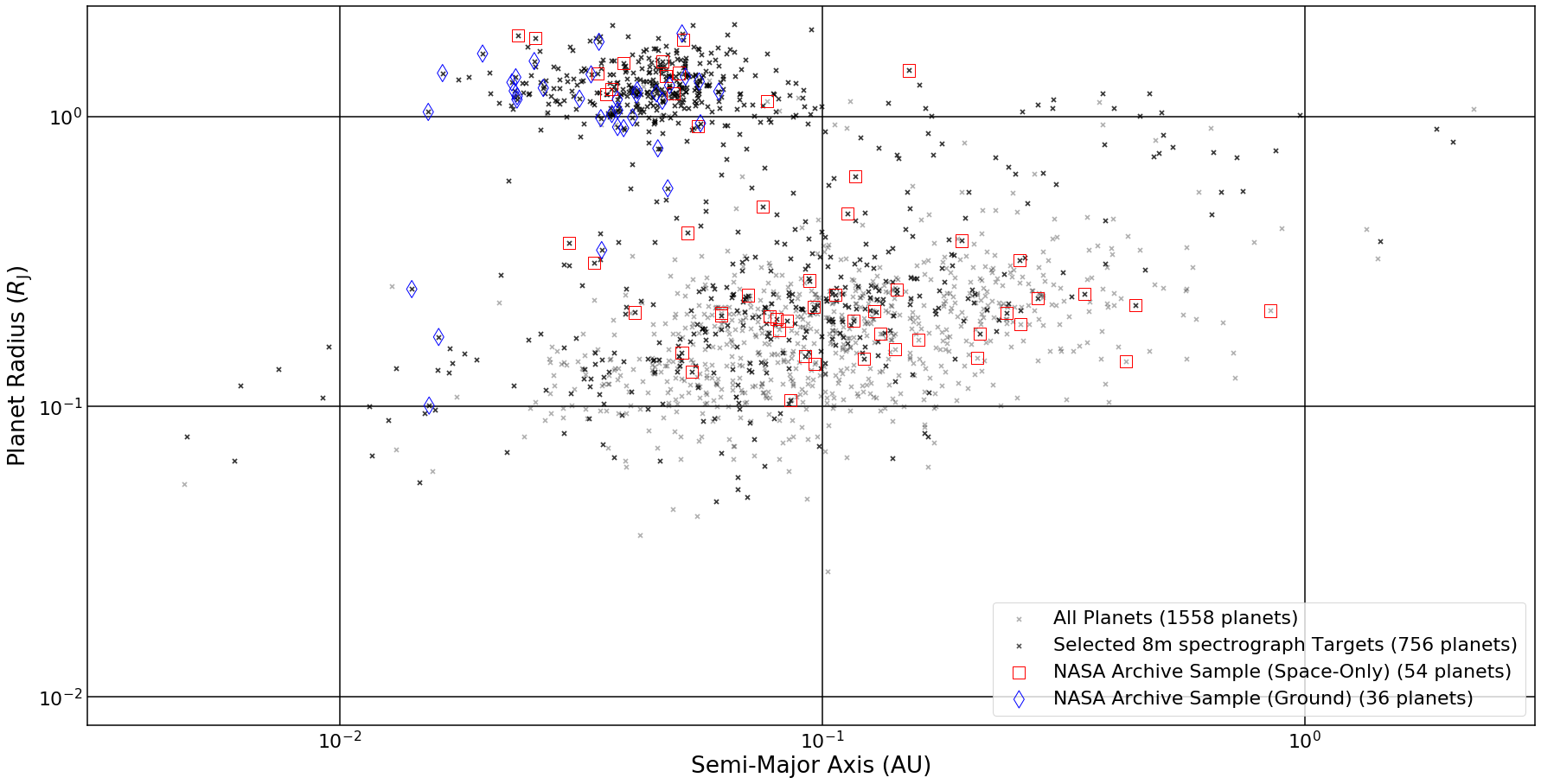}
	\includegraphics[width=0.77\textwidth]{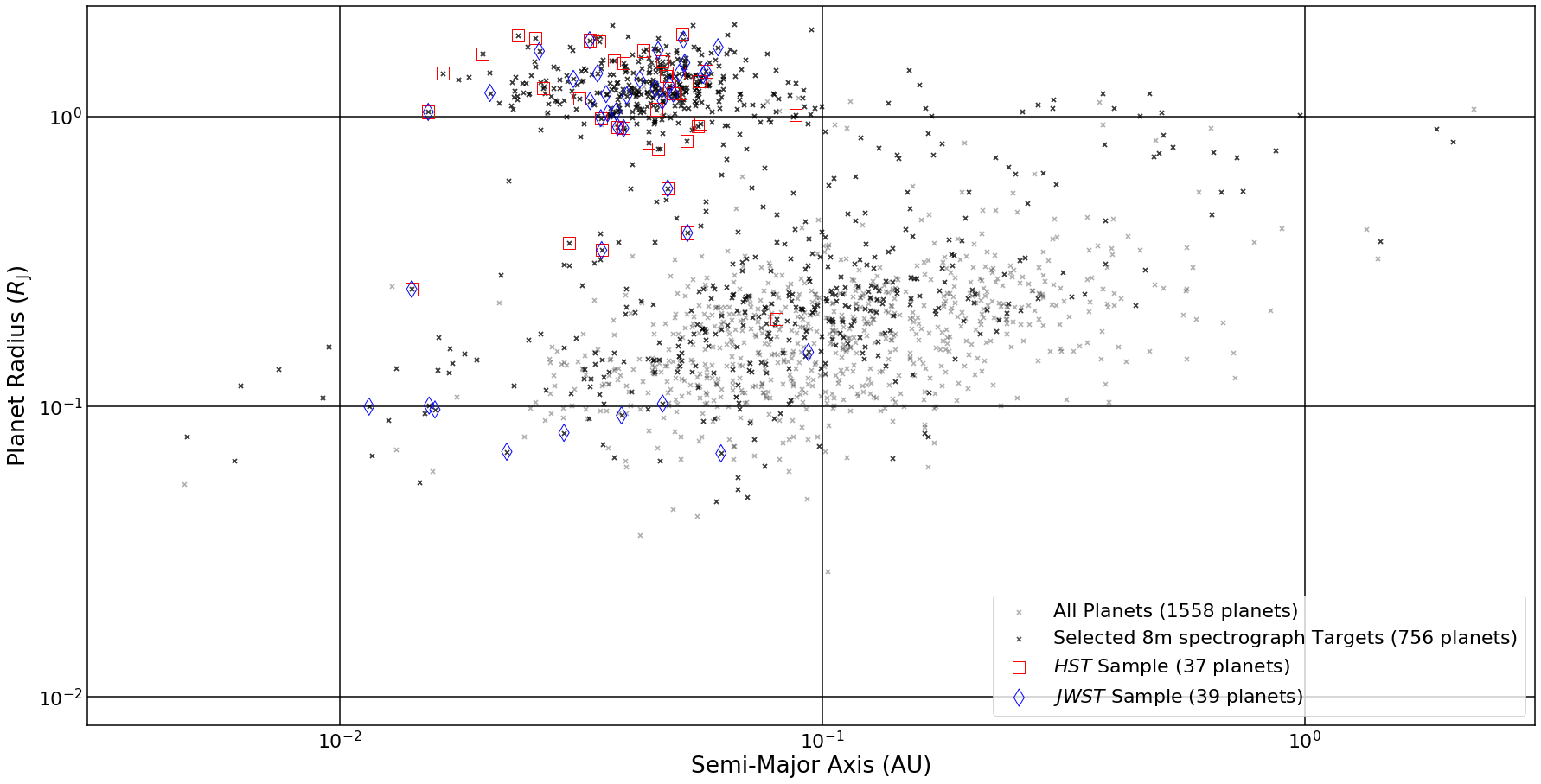}
	\caption{Cross-reference plots of our metric cut with a selection of pre-existing catalogues. In the first panel, green squares are the sample of 117 planets in the ATMO grid of forward models that have been identified as observationally significant. In the second panel, red squares and blue diamonds are planets present in the NASA Exoplanet Archive spectral database, with the former having no recorded observations from the ground present. This also includes the recent result of \protect\cite{WASP-107He}. In the third panel, red squares are planets which have been targeted by the \textit{HST} as selected by \protect\cite{FuHSTStats}, and also including the recent individual studies of \protect\cite{HAT-P-7HST}, \protect\cite{WASP-107Water}, \protect\cite{WASP-103HST} and \protect\cite{WASP-107He}. Blue diamonds are planets that have been identified as priority targets for the \textit{JWST}, collated from \protect\cite{JWSTSynth}, \protect\cite{JWSTComm}, \protect\cite{JWSTEarths} and \protect\cite{JWSTERS}.}
	\label{Sec4Combined}
\end{figure*}

The NASA Exoplanet Archive's Transmission Spectroscopy table\footnote{https://exoplanetarchive.ipac.caltech.edu/cgi-bin/TblView/nph-tblView?app=ExoTbls\&config=transitspec}  holds a volunteer-submitted list of successfully studied planets, including the instruments used and the wavelengths observed. Figure \ref{Sec4Combined} shows a number of planet samples, including those already studied and some which have been proposed for future study. Those that are selected by our decision metric for the 8m spectrograph are shown by black crosses. Planets present in the NASA Archive are over-plotted in two subsets, red squares for the space-based sample and blue diamonds for the ground-based sample. 

\begin{figure}
	\includegraphics[width=\columnwidth]{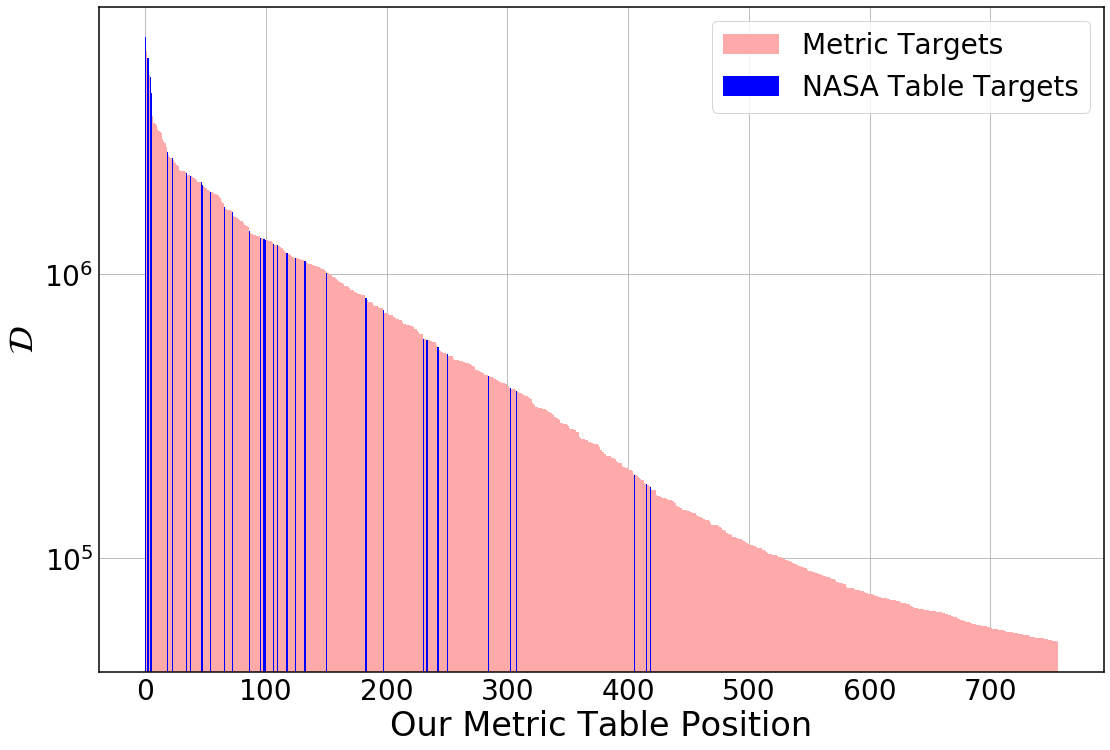}
	\caption{The metric versus expert selection; ranked table positions of previous targets successfully observed from the ground in the NASA Exoplanet Archive's Transmission Spectroscopy table, compared with their metric score from our 8m spectrograph example configuration. Previous targets have been observed using a range of ground-based facilities, but their clustering in rank position nonetheless correlates well with metric score, indicating that the scheme performs well as a means of selecting good transmission spectroscopy targets.}
	\label{NASAPreface1}
\end{figure}

It is apparent that all planets that are identified as having been studied previously from the ground (in some cases, in conjunction with space-based observations, shown by blue diamonds in Figure \ref{Sec4Combined}) are picked up by our decision metric (Eqs. \ref{OldMetric} and \ref{MetMass}). This sample is shown by the blue bars in Figure \ref{NASAPreface1}, which also shows the rank-ordered distribution of decision metric scores in pink. This confirms that our metric, despite its simplicity, captures enough relevant physics to reliably return sets of good targets.

The rank-ordered decision metric distribution in Figure~\ref{NASAPreface1} suggests two potential strategies for target selection. Firstly one can simply select targets starting from the target with highest metric score and then descending down the list. Alternatively, one can use the metric scores as weights and draw targets by a weighted random selection. In this case the rank ordered distribution represents a probability distribution function. Such an approach may be preferable if one is seeking to sample targets over a wider range of parameter space and avoid the possibility that the highest ranked targets may be driven by common characteristics.

\begin{figure}
	\includegraphics[width=\columnwidth]{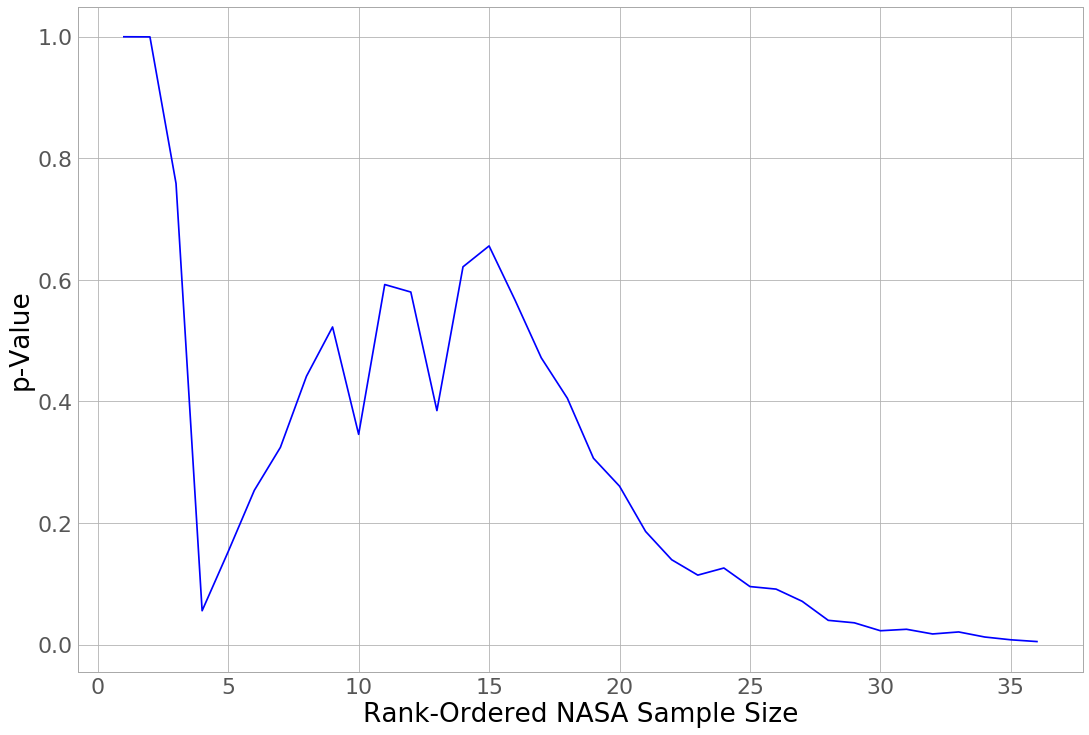}
	\caption{$p$-values of two-sample Kolmogorov-Smirnov (KS) tests, performed using the decision metric distribution and a varying subset of the NASA sample. The latter subset ranges from the lowest ranked up to some upper rank, which is adjusted to vary the number of included planets from 1 to 36 (the entire sample). The decision metric distribution comprises 419 planets with ranks between the lowest and highest NASA ranked planets. The change in $p$-value tracks how the two distributions appear similar or dissimilar as we include higher ranked planets. At the point where the sample includes 30 out of the 36 NASA planets, the $p$-value indicates that the hypothesis where the two samples are drawn from the same distribution can be ruled out with 95\% confidence.}
	\label{KSTest}
\end{figure}

Interestingly, Figure~\ref{NASAPreface1} suggests that previously observed targets show evidence of both approaches. Figure \ref{KSTest} shows the result of two-sample Kolmogorov-Smirnov (KS) testing, performed using the NASA sample and the rank-ordered decision metric distribution in Figure~\ref{NASAPreface1}. Starting from the lowest ranked planet in the NASA sample and cutting the sample at some higher rank position to contain some fixed number of NASA entries, we compute KS $p$-values as a function of the NASA sample size. The sample size is increased until it includes all of the highest ranking planets. Under a null hypothesis that both samples are drawn from the same underlying distribution, we should observe that the $p$-values do not drop below an acceptability threshold (e.g if $p$ drops below 0.01 then we can reject the hypothesis that the NASA and decision metric distributions are drawn from the same underlying distribution with at least 99\% confidence). In Figure~\ref{KSTest} we see that the $p$-value is generally maintained above 0.1 until the sample starts to include the 10 highest scoring NASA entries (i.e. sample sizes exceeding 27 NASA entries). The very highest ranking NASA targets appear to broadly follow a strategy of simply being selected by high ${\cal D}$ score, whilst for the large majority of the NASA sample they are consistent with being drawn by weighted selection.

Considering targets only observed from space with \textit{HST} and/or \textit{Spitzer} (marked with red squares in the second panel of Figure \ref{Sec4Combined}), the majority of these fall into the larger population of sub-Jupiter-radius planets. Many of these are \textit{Kepler} planets in orbit around faint host stars, with nearly all of these coming from the study of \cite{DesertSpitzer}, which employed \textit{Spitzer} to estimate false positive rates from the \textit{Kepler} satellite and find possible blended stars in their data set, not to identify targets for atmospheric study. Under the broad cut taken for the purposes of comparison with other instruments, our metric does select some of these targets; however, their $\cal D$ scores are generally low, and many would be excluded by a cut intended for observing purposes.

The red square markers in the third panel of Figure \ref{Sec4Combined} show the \textit{HST} targets selected by \cite{FuHSTStats}, as well as a selection of subsequent recent studies. These planets were initially discovered by ground-based surveys (Table 1 of \citealt{FuHSTStats}), and as such, all of them are also picked up by our metric as being top targets. The blue diamonds in the third panel of Figure \ref{Sec4Combined} represent a sample of planets identified as priority targets for the upcoming \textit{JWST}. These are drawn from planets identified by \cite{JWSTSynth}, \cite{JWSTComm} and \cite{JWSTEarths}, with the latter looking at the potential of the \textit{JWST} for studying Earth-like planets. In particular, the TRAPPIST-1 system of seven rocky planets (\citealt{TRAPPISTDisc}) is identified as a high-priority target. \cite{JWSTERS} name WASP-18b, -43b and -79b as targets for the ERS program, drawn up for a spring 2019 launch.

All of the planets identified by \cite{JWSTSynth} and \cite{JWSTComm} are selected by our metric, along with the three named ERS targets. WASP-76b is flagged as a previously-unstudied target (\citealt{WASP-76bDisc}). As we might expect, the ground-based 8m spectrograph cannot effectively target all of the Earth-like planets identified by \cite{JWSTEarths}, which is to be expected; the authors find that the \textit{JWST} will need to capture multiple transits in order to reach 5$\sigma$-detection levels, but this is well within reach of an extended campaign. Although all of the targets are selected in this broad cut, in practice many transits would need to be captured in order to attain a useful $\cal S/N$. A significant fraction of these targets would be excluded under a tighter observationally-motivated cut, but not all, underscoring the continuing utility of large ground-based facilities in the era of $JWST$.

The green squares in the first panel of Figure \ref{Sec4Combined} make up a larger theoretical sample; that chosen by \cite{ATMO} in the construction of their grid of forward spectral models, using the ATMO 1D-atmosphere modelling suite. Their selection is based on the expected signal of an atmosphere of one scale height, with the scaling

\begin{equation}
\sn_{\rm ATMO} \propto 10^{-0.2 m_* (V)}  \frac{R_{\rm p} H}{R_*^2} ,
\label{ATMOScaling}
\end{equation}

\noindent which was also arrived at independently by \cite{ZellemMetric}. For further comparison, another similar metric intended for use with the \textit{JWST} was recently published by \cite{KemptonMetric}, taking the following form

\begin{equation}
{\rm TSM} \propto S_{\rm JWST} 10^{-0.2 m_* (J)} \frac{R_{\rm p}^3 T_{\rm eq}}{M_{\rm p}R_*^2},
\label{KemptonScaling}
\end{equation}

\noindent where $S_{\rm JWST}$ is a simulation-dependent scaling factor. This is in comparison to our overall metric scaling of

\begin{equation}
{\cal D} \propto 10^{-0.2 m_* (\lambda)}  t_{14}^{1/2}   \left(\frac{\texp}{\texp + \tov}\right)  \frac{R_{\rm p}H}{R_*^2} .
\label{DScaling}
\end{equation}

Using Eq. (\ref{ATMOScaling}), the authors construct contours of $\sn$ based around the planet WASP-12b, which tops their ranked list. For a cut-off, WASP-43b is used, as it falls on the contour of $\sn_{\rm ATMO}$=1 in the $V$-band (see Figure 2 of \citealt{ATMO}).

\begin{figure*}
	\includegraphics[width=0.7\textwidth]{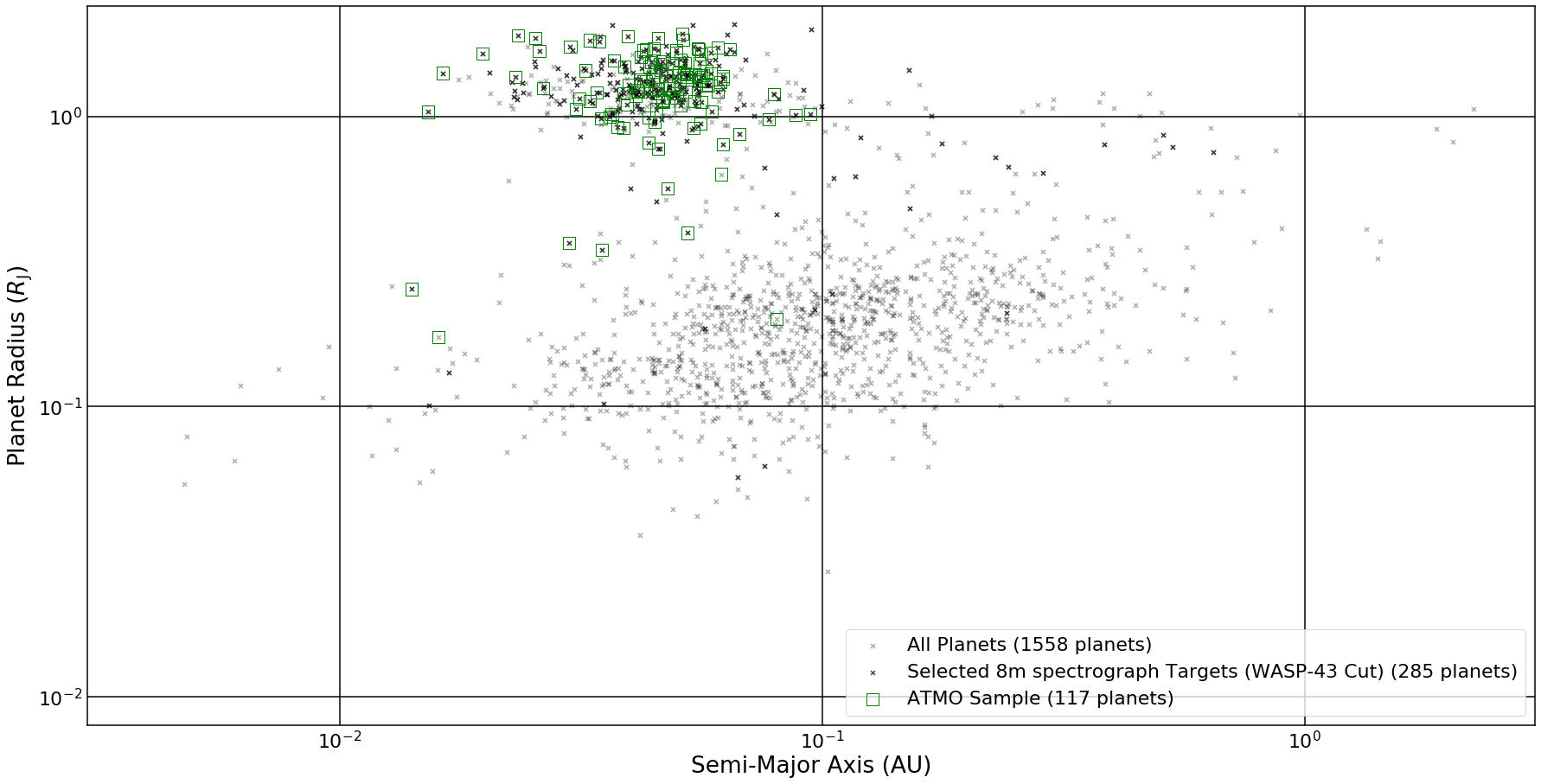}
	\includegraphics[width=0.7\textwidth]{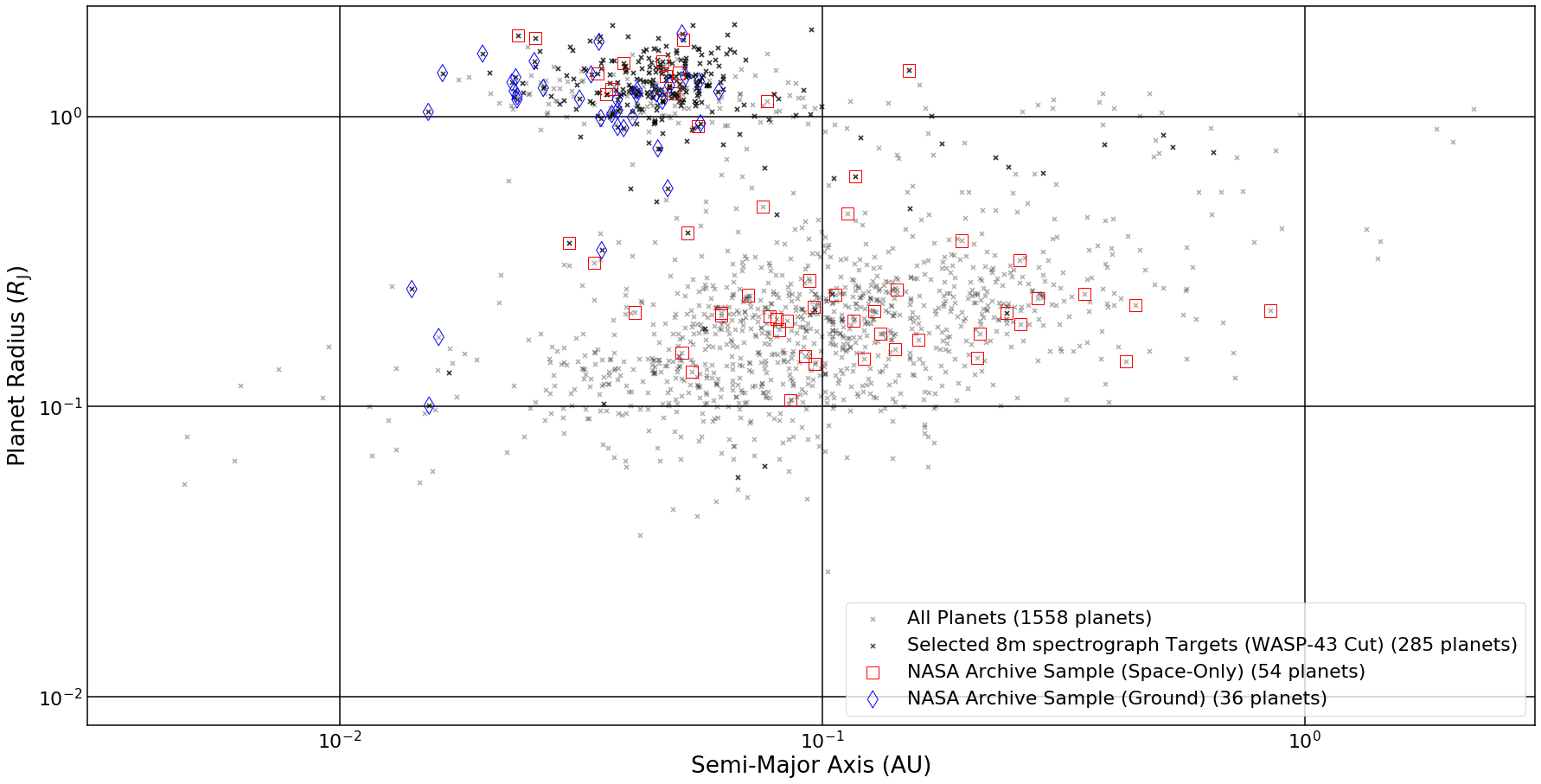}
	\includegraphics[width=0.7\textwidth]{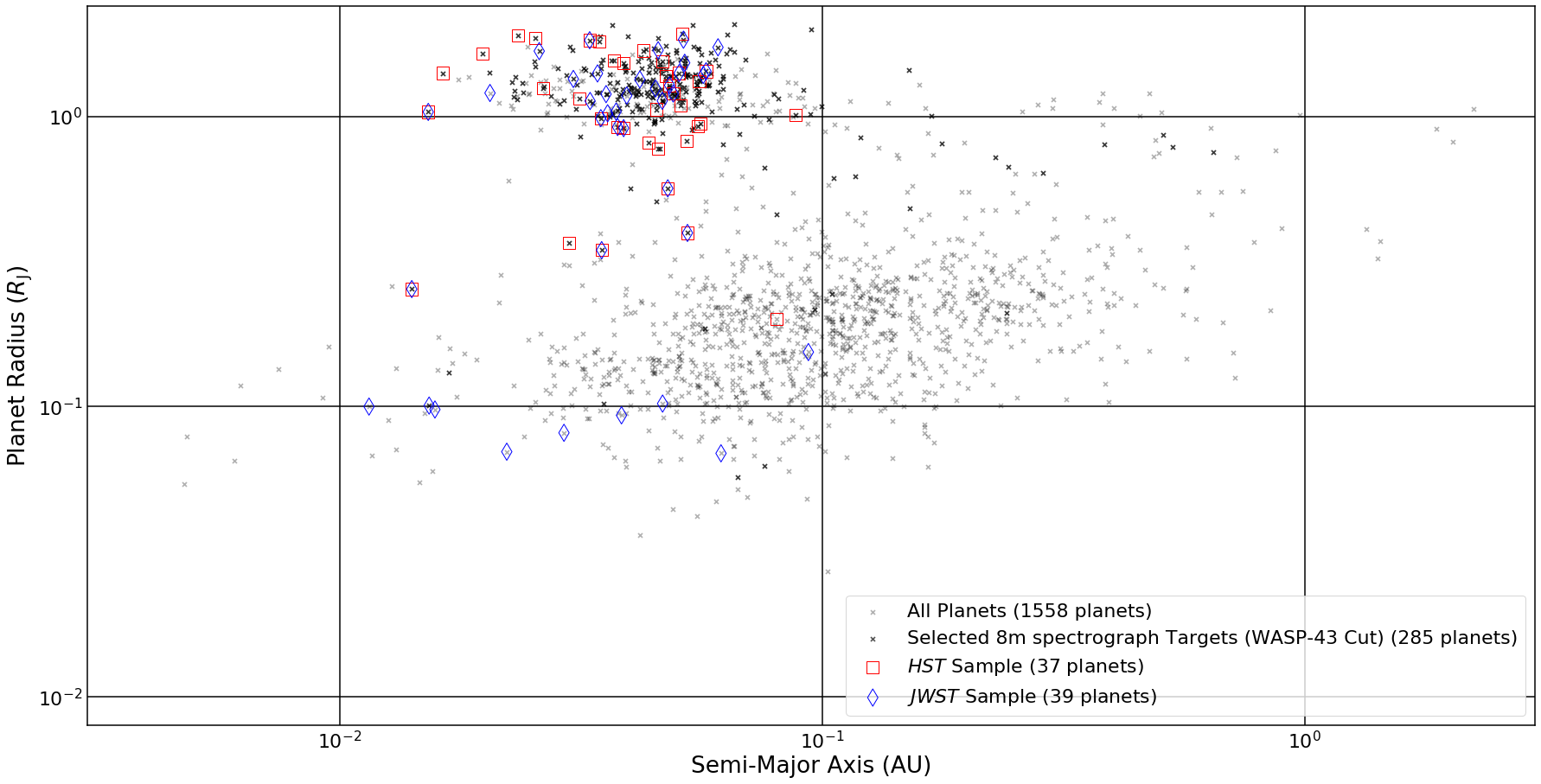}
	\caption{As Figure \ref{Sec4Combined}, but restricting the metric sample of metric-selected targets such that WASP-43b is the cut-off point.}
	\label{Sec4CombinedWASPCut}
\end{figure*}

\begin{figure}
	\includegraphics[width=\columnwidth]{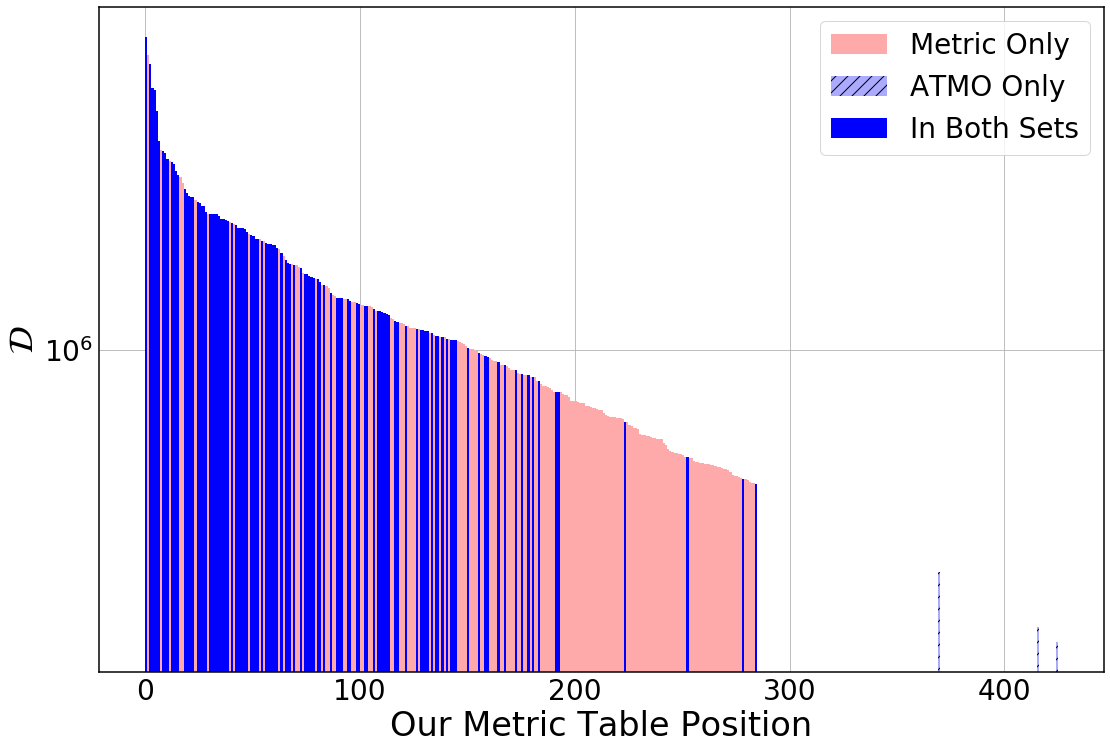}
	\caption{Bar plot of the metric score $\cal D$ versus table position, for a metric cut taken at WASP-43b for both samples. Three planets are identified by ATMO but not our metric.}
	\label{EitherOrBothPlot}
\end{figure}

Inspecting Figure \ref{Sec4Combined}, we see complete agreement between our metric cut and the ATMO sample. Figure \ref{Sec4CombinedWASPCut} extends this inspection, by instead taking the metric cut at WASP-43b (rank position 285, corresponding to a 80\% cut in the total cumulative distribution) to better compare the two samples. In doing this, a divergence occurs, as illustrated in Figure \ref{EitherOrBothPlot}. 114 planets are identified by both our metric and ATMO, meaning there exists a small set of planets identified for the ATMO sample but not our metric, and a significant population identified by our scheme and not for ATMO.

Referring to TEPCat, we find that many of the planets flagged only by our metric have had no previous study done on them; i.e. the only reference listed is their discovery paper. A significant fraction of these planets were only announced in 2017 and 2018; their not being included in ATMO is likely because they were simply unknown to science at the time. We note that all planets in the ATMO sample have known masses, so $\cal D_{\rm mass}$ is employed in all cases, yielding only three planets identified by ATMO but not our metric. These are the super-Earth targets HD 97658b (\citealt{HD_97658b_Disc}) and 55 Cancri e (\citealt{55_Cnc_Disc1}, \citealt{55_Cnc_Disc2}), and WASP-86b (\citealt{WASP-86b_Disc1}, also announced as KELT-12b by \citealt{WASP-86b_Disc2}).

To explain these discrepancies, we must consider the different scalings of the two metric schemes. Our metric has a stronger dependence on planet radius, as we saw in Section \ref{bias}, but this alone cannot entirely explain the disparity between the planets identified by our metric and those for ATMO. Compared to the scaling employed by \cite{ATMO}, given in Eq. (\ref{ATMOScaling}), $\cal D$ in Eq. (\ref{DScaling}) has additional dependencies on transit duration, depth and the observational efficiency of the telescope being used. These correspond to the second, third and fourth terms of Eq. (\ref{DScaling}) respectively.

In order to make the metric cut, a planet must perform well in all of these aspects, with its rank position being suppressed if it fares poorly in any one of them. Of the additional terms in Eq. (\ref{DScaling}), the transit depth term will be approximately unity, even for hot Jupiters producing the deepest transits. Therefore, it is the action of the transit duration and efficiency terms which drives the disparity between our metric selection and those planets identified for ATMO. We have seen that metric will disfavour those planets with short transit durations, or planets for which $t_{\rm exp} \ll t_{\rm over}$. 

Both HD 97658b and 55 Cancri e orbit bright host stars ($V$ = 7.7 and 5.95 respectively). For this case, they are in the regime where $t_{\rm exp} \ll t_{\rm over}$, and their $\cal D$ scores are suppressed as a result. The discrepancy for WASP-86b/KELT-12b stems from two sets of planet parameters in tension with one another - \cite{WASP-86b_Disc1} identify the planet as a dense sub-Jupiter of $M_{\rm p}\ = 0.82 \pm 0.06 M_{\rm J}, R_{\rm p}\ = 0.63 \pm 0.01 R_{\rm J}$, while \cite{WASP-86b_Disc2} announce a bloated planet of $M_{\rm p}\ = 0.95 \pm 0.14 M_{\rm J}, R_{\rm p}\ = 1.79^{+0.18}_{-0.17} R_{\rm J}$. The TEPCat catalogue used by our metric has the former set of values listed, while \cite{ATMO} have adopted the latter.

Our metric is also restricted to observing a single transit for each planet (Eq. \ref{OldMetric}); planets not identified by the metric (operating in single-pass mode) will therefore need extended campaigns to observe effectively. In the regime of limited follow-up resources (as is expected to be the case in the near future), these should be considered lower-priority targets, particularly in cases where archived spectra already exist.

\section{Conclusions}
\label{Conclude}

The SPEARNET team is undertaking a transmission spectroscopy survey of gas and ice giant planets using a globally-distributed heterogeneous telescope network. Our aim is to use objective criteria for the selection of targets in order to facilitate population studies of exoplanet atmospheres. We are employing a three-stage approach involving: i) short-listing targets observable from each telescope asset; ii) using a \textit{decision metric} to optimally pair observable exoplanet targets with available telescope assets; iii) an assessment of how well new data improves characterisation of the target atmosphere. In this paper we have described and validated our general framework for Stage~(ii). A future paper will describe Stage~(iii) and how it feeds back to inform and update Stage~(ii).

We have argued for the imminent need for a decision metric in order to efficiently choose samples of exoplanets in the new era of comparative exoplanetology. Meeting the challenges of this new era will involve handling the transition from a target-starved to a telescope-starved regime. We argue that telescopes with a range of apertures and capability will have an important role to play in the follow-up of the vast numbers of nearby planets that will be discovered by new generation wide-area surveys such as \textit{NGTS}, \textit{TESS} and \textit{PLATO}. It is therefore crucial to have an appropriate and efficient means to allocate targets to observing resources in order to maximise science returns.

We have presented a decision metric that is designed to select the most promising targets and to pair the most suitable targets to the most suitable observing facility. The metric can be used to assess the relative performance of different telescopes for transmission spectroscopy studies. The ability to select targets objectively also facilitates population studies. One form of our decision metric ${\cal D}$ relies only on primary transit observables, whilst another form ${\cal D}_{\rm mass}$ also includes knowledge of the planet mass that may come from follow-up observations.

If planet masses are already known then our mass-sensitive metric, ${\cal D}_{\rm mass}$, is the preferred statistic. However it will likely take considerable telescope resources and time to obtain masses for the majority of the candidates that will come from the new generation surveys. This presents a major potential bottle-neck to the progress of atmosphere studies. However, we show that in the absence of planet mass information ${\cal D}$ forms a credible proxy to ${\cal D}_{\rm mass}$. We validate our metric by showing that it tends to highly rank samples of previously successfully studied systems.

\section*{Acknowledgements}

JSM and JH are supported by PhD studentships from the Science and Technology Facilities Council. EK and IM are supported by the United Kingdom's Science and Technologies Facilities Council (STFC) grant ST/P000649/1. This work is also supported by a National Astronomical Research Institute of Thailand (NARIT) research grant.

We thank the anonymous referee for several useful suggestions that have significantly improved the paper. 

This research has made use of the SVO Filter Profile Service (http://svo2.cab.inta-csic.es/theory/fps/) supported from the Spanish MINECO through grant AyA2014-55216, the Transiting Exoplanet catalogue hosted by Keele University and the NASA Exoplanet Archive, which is operated by the California Institute of Technology, under contract with the National Aeronautics and Space Administration under the Exoplanet Exploration Program. 




\bibliographystyle{mnras}
\bibliography{MetricPaperRefs} 


\bsp	
\label{lastpage}
\end{document}